\documentclass{elsart}
\usepackage{graphics}
\usepackage{graphicx}
\usepackage{lscape}
\usepackage{amssymb}
\usepackage{latexsym}
\usepackage{verbatim}
\usepackage{multirow}
\journal{Chemical Physics}
\begin{document}

\begin{frontmatter}

\title{Time\textendash dependent density functional study of the electronic 
spectra of oligoacenes in the charge states -1, 0, +1, and +2}

\author[label1]{G.~Malloci\corauthref{cor}},
\corauth[cor]{Corresponding author. Tel: +39\textendash070\textendash675\textendash4915;
Fax: +39\textendash070\textendash510171} 
\ead{gmalloci@ca.astro.it}
\author[label1]{G.~Mulas},
\author[label2,label1]{G.~Cappellini},
\author[label3]{C.~Joblin}

\address[label1]{INAF \textendash{} Osservatorio Astronomico 
di Cagliari\textendash Astrochemistry Group, Strada 54, Localit\`a Poggio dei Pini, 
I\textendash09012 Capoterra (CA), Italy}
\address[label2]{CNR\textendash Sardinian Laboratory for Computational Materials
Science, CNISM and Dipartimento di Fisica, Universit\`a degli Studi di Cagliari,
Cittadella Universitaria, Strada Prov.~le Monserrato\textendash Sestu Km~0.700,
I\textendash09042 Monserrato (CA), Italy}
\address[label3]{Centre d'Etude Spatiale des Rayonnements \textendash{}
Universit\'e Toulouse~3 \textendash{} CNRS \textendash{} Observatoire Midi\textendash Pyr\'en\'ees,
9 Avenue du Colonel Roche, 31028 Toulouse Cedex 4, France}
\begin{abstract}
We present a systematic theoretical study of the five smallest oligoacenes 
(naphthalene, anthracene, tetracene, pentacene, and hexacene) in their anionic,
neutral, cationic, and dicationic charge states. We used
density functional theory (DFT) to obtain the ground\textendash state 
optimised geometries, and time\textendash dependent DFT (TD\textendash DFT) to evaluate 
the electronic absorption spectra. Total\textendash energy differences enabled us
to evaluate the electron affinities and first and second ionisation 
energies, 
the quasiparticle correction to the 
HOMO\textendash LUMO energy gap and an estimate of the excitonic 
effects in the neutral molecules.
Electronic absorption spectra have been 
computed by combining two different implementations of TD\textendash DFT: the 
frequency\textendash space method to study general trends as a function of charge\textendash state 
and molecular size for the lowest\textendash lying in\textendash plane long\textendash polarised and 
short\textendash polarised { $\pi\to\pi^\star$} electronic transitions,
and the real\textendash time propagation scheme to obtain the whole photo\textendash absorption 
cross\textendash section up to the far\textendash UV.
Doubly\textendash ionised PAHs are found to display strong electronic transitions of 
$\pi\to\pi^\star$ character in the near\textendash IR, visible, and near\textendash UV spectral ranges, like 
their singly\textendash charged counterparts. While, as expected, the broad 
{ plasmon\textendash like} structure with its maximum at { about} 17\textendash18 eV is 
relatively insensitive to the charge\textendash state of the molecule, a systematic 
decrease with increasing positive charge of the absorption cross\textendash section 
between $\sim$6 and $\sim$12~eV is observed for each member of the class.
\end{abstract}


\end{frontmatter}

\section{Introduction}
\label{introduction}

Polycyclic aromatic hydrocarbons \cite{clar} (PAHs) are a large class of 
conjugated $\pi$\textendash electron systems of fundamental importance in many research 
areas of chemistry as well as in astrophysics and materials science. The carbon
skeletons of PAHs may be considered as small pieces of graphite planes and,
as such, they have been proposed as precursors to extended carbon networks 
such as fullerenes and carbon nanotubes \cite{ful}.
PAHs are of high interest in environmental chemistry due to their 
carcinogenicity and their ubiquity as air pollutants produced by the 
combustion of organic matter \cite{environment}. In the astrophysical context,
PAHs are found in carbonaceous meteorites \cite{hah88} and in interplanetary 
dust particles \cite{cle93}. Based on the astronomical observation of IR,
visible and UV spectroscopic features, neutral and charged PAHs are 
thought to be the most abundant molecules in space after molecular hydrogen 
and carbon monoxide \cite{tie05}.

Oligoacenes, or simply acenes, are a subclass of catacondensed PAHs (with all 
carbon atoms on the periphery of the ring system) consisting of fused benzene 
rings joined in a linear arrangement. In their crystalline state these organic 
semiconducting materials have received particular attention in the field of 
electronics and photonics \cite{solids}. Acenes and their derivatives are 
being increasingly used as active elements in a variety of opto\textendash electronic 
devices such as organic thin\textendash film field\textendash effect transistors \cite{nel98kla02}, 
light\textendash emitting diodes \cite{azi02kan04}, photovoltaic cells \cite{sen02yoo04},
and liquid crystals \cite{shi03}. Organic electronics based on functionalised 
acenes and heteroacenes is presently a very active field of research
\cite{hetero}.

Since the early work by Clar \cite{clar} and Platt \cite{platt}, 
there has been a wide\textendash ranging interest in the electronic properties of PAHs 
using different spectroscopic techniques, such as absorption \cite{abs}, 
electron\textendash energy\textendash loss \cite{eels}, fluorescence \cite{fluor}, polarisation
\cite{mcd}, photoelectron \cite{pe}, and photoion mass spectrometry \cite{joc}.
Thanks to these studies, the electronic spectra of neutral PAHs are known to 
be composed { of} two main regions: (i) the broad { plasmon\textendash like} 
excitation peaking at $\sim$17\textendash18~eV, which involves $\pi\to\sigma^\star$, $\sigma\to\pi^\star$, $\sigma\to\sigma^\star$, 
and Rydberg spectral transitions, and (ii) the single\textendash particle excitation 
part below a few eV, where the lowest energy singlet\textendash singlet $\pi\to\pi^\star$ 
transitions occur. The four lowest transitions of neutral PAHs are usually 
described by the Clars's notation $p$, $\alpha$, $\beta$, $\beta^\prime$ \cite{cla41}, or $^1L_a$, 
$^1L_b$, $^1B_b$, $^1B_a$ according to the empirical model of Platt \cite{pla49}. In 
catacondensed PAHs these transitions are characterised by the following 
intensities and oscillator strengths: $p$, weak, $f\approx0.01\textendash0.1$; $\alpha$, 
very weak, $f\approx0.001$; $\beta$, very strong, $f\approx1.0$; $\beta^\prime$, medium strong, 
$f\approx0.1\textendash1.0$ (e.g., \cite{joc97}). { In the case of oligoacenes, in 
particular, the transition dipole moment for the $p$\textendash band lies 
along the short\textendash axis of the molecule, while for the $\alpha$ and $\beta$\textendash bands it
lies along the long\textendash axis. In the following we 
will refer to these transitions as ``short\textendash polarised'' and 
``long\textendash polarised'', respectively.}

Charged PAHs have been the subject of extensive spectroscopic studies in frozen
glassy organic solids \cite{shida}. These experiments showed that PAH radical 
cations and anions: (i) display intense optical transitions at lower energies 
than their parent molecule, and (ii) have very similar electronic spectra, in 
qualitative agreement with the particle\textendash hole equivalence in the pairing 
theorem of H\"uckel's theory \cite{shi73}. Currently, the interest on charged 
PAHs comes mainly from basic research in astrophysics because PAHs are 
expected to exist in space in different charge states depending on the physical
conditions (e.g., UV flux, electron density, etc.) of the host environment 
\cite{tie05}. This has motivated a large amount of laboratory work based first 
on matrix isolation spectroscopy \cite{mis} and, more recently, laser mass 
spectroscopy \cite{brechi}, spectroscopic studies of the molecules trapped in 
helium droplets \cite{helium}, and a high sensivity photo\textendash absorption technique
in free jets called cavity ring\textendash down spectroscopy \cite{crd}. 

\begin{figure}
\begin{center}
\includegraphics*[width=2cm]{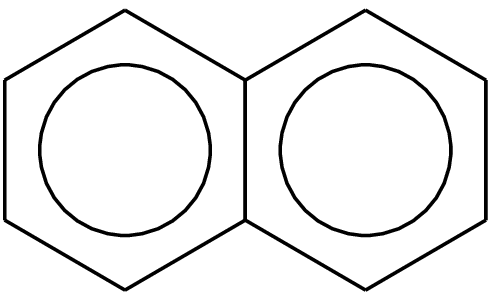}\\
\includegraphics*[width=3cm]{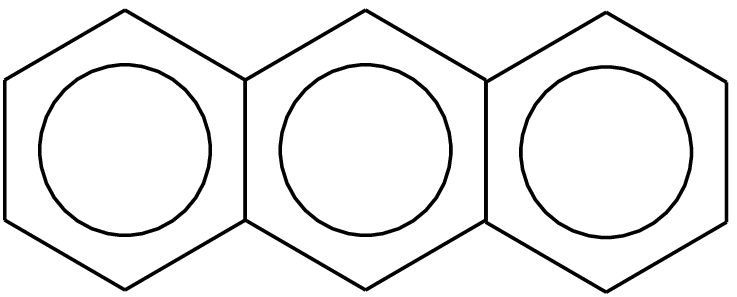}\\
\includegraphics*[width=4cm]{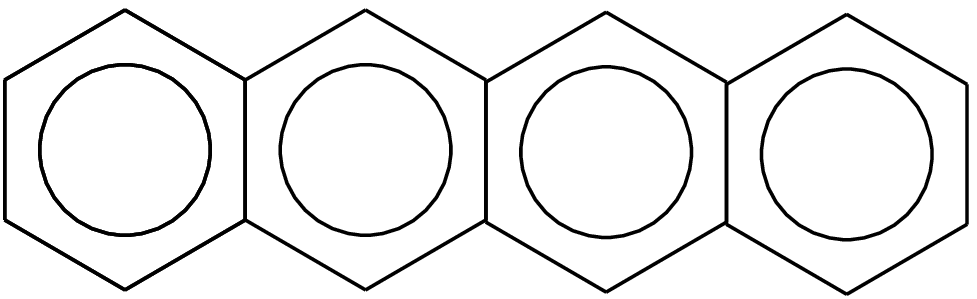}\\
\includegraphics*[width=5cm]{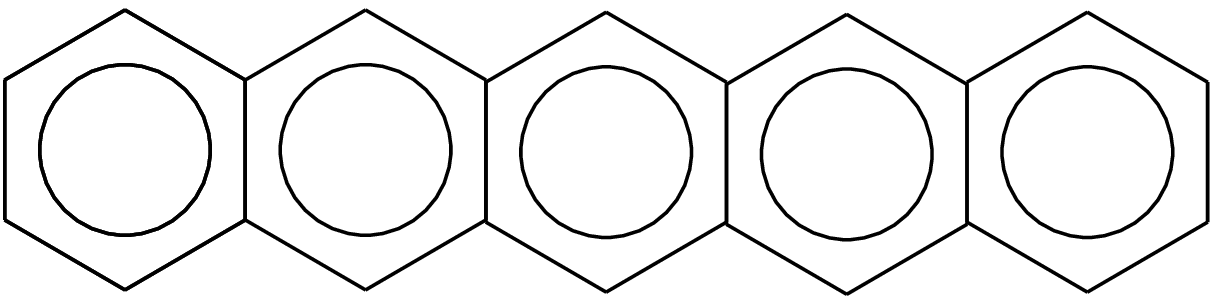}\\
\includegraphics*[width=6cm]{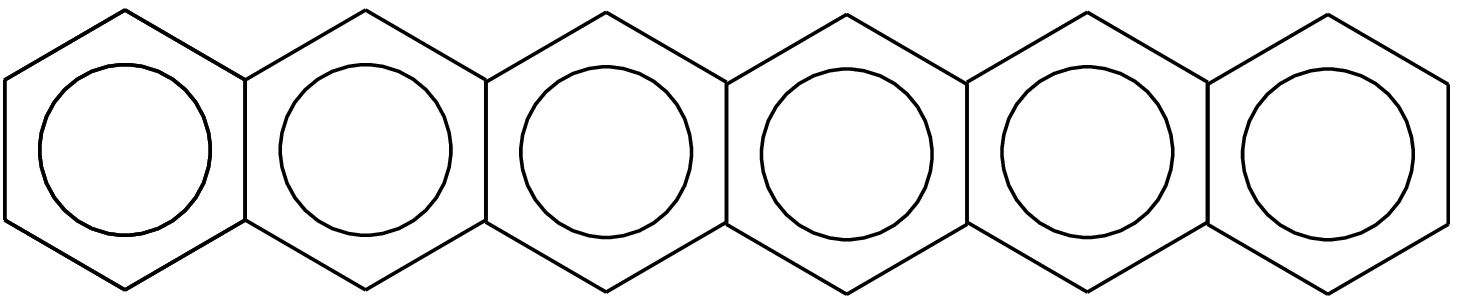}\\
\end{center}
\caption{Oriented geometries of the molecules considered. From top to bottom: 
naphthalene, anthracene, tetracene, pentacene, and hexacene. The molecules are 
supposed lying in the $x\textendash y$ plane, the $y$ axis being the longer one.}
\label{geoms}
\end{figure}

From an astrophysical point of view the knowledge of the electronic absorption 
spectra of PAHs in all their relevant charge states is of fundamental 
importance for our understanding of their photophysics in space. While this 
concerns the whole energetic range excitable in a typical interstellar 
environment, i.e., from the visible to the far\textendash UV, very few experimental
data are available for charged PAHs in this spectral range due to the 
limitations which are intrinsic to the laboratory techniques more widely used. 
As a part of a more extensive study \cite{mal} towards the knowledge of the 
spectral properties of a large sample of PAHs to be modelled in astrophysical 
environments \cite{mul}, we report in this paper a detailed study of the 
electronic absorption spectra of 
the five smallest oligoacenes naphthalene, anthracene, tetracene, pentacene, 
and hexacene in the charge states most relevant for astrophysical applications,
i.e., -1, 0, +1, and +2. The geometries of the molecules considered are
sketched in Fig.~\ref{geoms}. The theoretical methods we used for both
ground\textendash state and excited\textendash state calculations have been validated for the basic
aromatic unit benzene in its neutral form, for which a large amount of
spectroscopic data are available.

There have been many systematic studies of benzenoid hydrocarbons, including
acenes, using quantum\textendash chemical calculations to obtain, e.g., heats of 
formation \cite{pec90}, infrared spectra \cite{def93}, and C\textendash H bond 
dissociation energies \cite{cio96}. While the electronic properties of neutral 
acenes in the near\textendash IR, visible, and near\textendash UV spectral ranges are thoroughly 
characterised up to the sizes of hexacene \cite{uptohexa,uptohexa1} and some 
studies have been done for larger acenes \cite{larger}, a comparatively smaller
amount of work exists for their corresponding mono\textendash cations \cite{cations} and 
mono\textendash anions \cite{hal00,hir03,anions}. To the best of our knowledge, however, 
the electronic spectra of charged PAHs in the far\textendash UV spectral domain 
have not been measured to date.
In addition, a detailed study of the electronic 
excitation properties of PAHs in their doubly\textendash ionised state has been missing 
until now. First proposed twenty years ago \cite{dications}, the possible 
presence of PAH dications in the interstellar medium has recently received 
further support based on the proposition that these molecules could contribute 
to the luminescence observed in the red part of the visible spectrum in many 
interstellar sources \cite{wit06}. While laboratory measurements of the yield 
of optical fluorescence and phosphorescence by these species are needed to test
this hypothesis, the determination of their electronic absorption spectra is 
motivated.

We used density functional theory (DFT) \cite{dft_papers} and its 
time\textendash dependent extension (TD\textendash DFT) \cite{tddft}, which are methods of choice 
for this type of investigations for large molecules. { Since excitation 
energies and oscillator strengths within TD\textendash DFT can be computed following two 
different strategies (see Sect.~\ref{td_dft} for details), we used both of them
to obtain: (i) the whole absolute photo\textendash absorption cross\textendash sections up to the 
soft x\textendash ray region near 30~eV, and (ii) the positions, the oscillator 
strengths, and the leading configurations of the lowest\textendash lying valence
$\pi \to \pi^\star$ short\textendash polarised and long\textendash polarised electronic transitions, falling 
in the visible/near\textendash UV region, as a function of molecular size and charge 
state. In case (i), due to its numerical stability, we used the adiabatic 
local\textendash density approximation in the parametrisation of Perdew \& Zunger 
\cite{per81}, which has proven to yield reliable results for the dynamical 
polarisability of conjugated molecules \cite{mar07,yab99}. In case (ii) we 
used the hybrid B3LYP functional \cite{bec93} and the gradient\textendash corrected BLYP 
functional \cite{bec88,lee88}, which are widely used in the study of PAHs to 
obtain a large number of
molecular parameters, such as structures and energetics \cite{wib97,mar96}, 
vibrational spectra \cite{mar96,lan96,bau97}, ionisation spectra 
\cite{del01,del03}, ionisation energies \cite{mal07a,wib97,sch01,kad06}, 
electron affinities \cite{mal05,des00,rie01,mod06}, 
electronic excitations \cite{hir99,ngu02,par03,gri03,hir03,bau05,kad06},
the vibronic structure of absorption spectra \cite{die04a,die04b}, hydrogen
dissociation \cite{jol05,pin07}, and Jahn\textendash Teller effects 
\cite{kat99,kat01,kat02a,kat02b,kat05}.}

Although the approximate exchange\textendash correlation functionals we used have been 
developed for the electronic ground\textendash state, they are also routinely employed in
TD\textendash DFT calculations, and their application usually yields accurate results for
{ low\textendash lying} valence\textendash excited states { of both closed\textendash shell 
\cite{bau96,wib98} and open\textendash shell \cite{hir99b,hir99c} species}. { It is 
known, however, that these functionals show the wrong asymptotic behaviour, 
decaying faster than $1/R$ (i.e., exponentially) for large distances $R$ from 
the nuclei.} Among the well known and documented limitations of these methods 
\cite{dre05,bur05} { are: (i) the correct description of Rydberg 
\cite{rydberg}, doubly\textendash excited \cite{doubly}, and charge\textendash transfer \cite{ct} 
excited\textendash states, (ii)  the failure for large, extended $\pi$\textendash systems such as 
polyacetylene fragments and oligoporphyrins \cite{cai02}}, and (iii) the 
system\textendash size\textendash dependent errors found for the lowest short\textendash polarised excited 
states ($p$\textendash bands in Clar's notation) of neutral oligoacenes \cite{gri03}. 
{ Furthermore, the oscillator strengths computed by TD\textendash DFT are considered
to be only in qualitative agreement with experiments. In a rigorous assessment 
of the quality of TD\textendash DFT molecular oscillator strengths, for example, they 
were found to be reasonable but not in quantitative agreement with reliable 
experimental and theoretical values for small molecules such as CO, N$_2$, and 
CH$_2$O \cite{cas00}.}

In spite of { the above\textendash mentioned failures of the level of theory} we used,
it is known to be sufficient to identify the most intense { lowest\textendash lying} 
electronic excitations of neutral PAHs \cite{hei00,par03} as 
well as radical ions \cite{hir99,hal00,hir03,anions,series},  { which are 
found to match closely with the available experimental data both in terms of 
positions and intensity ratios. Despite the non\textendash physical exponential fall\textendash off 
of the exchange\textendash correlation functional used, in fact, these states all involve
excitations to and from delocalised, valence $\pi$\textendash orbitals, which
are not significantly affected by the shape of the 
exchange\textendash correlation potentials in the asymptotic region \cite{hir03}.} 
In particular, it has been shown that interesting trends exist in the vertical 
excitation energies and the oscillator strengths for homologous series of PAHs 
\cite{hal03,wei03,wei05}. { For the transitions with the largest 
oscillator strength in the oligorylenes perylene, terrylene, and quaterrylene, 
for example, a net increase of the oscillator strength per unit mass of carbon 
along the series has been found \cite{hal03}. This result might have important 
implications in astrophysics with respect to the long\textendash standing unsolved  
problem of the diffuse interstellar bands, about 300 unidentified absorption 
features observed in the near\textendash UV, visible, and near\textendash IR spectra of stars 
obscured by interstellar dust \cite{sar06}.}

Moreover, we used the so\textendash called delta\textendash self\textendash consistent\textendash field ({ $\Delta$SCF}) 
approach \cite{jon89}, evaluating total energy 
differences between the self\textendash consistent field calculations performed for the 
neutral and charged systems to obtain: (i) the vertical and adiabatic electron 
affinities and the first and second ionisation energies; (ii) the quasiparticle
correction (quasiparticle energies are associated with the addition or removal
of an electron) to the highest occupied molecular { orbital} (HOMO) \textendash{} 
lowest unoccupied molecular orbital (LUMO) gap. This quantity is related to 
molecular hardness, the { analogue} of the band gap of solids, defined as 
half the difference
between the ionization potential and the electron affinity, which is a key 
property characterising the chemical behaviour and reactivity of a molecule.
The comparison between the optical gap, i.e., the lowest singlet\textendash singlet 
excitation energy obtained with TD\textendash DFT, and the quasiparticle 
corrected HOMO\textendash LUMO gap enabled us to estimate the excitonic effects (due to 
the electron\textendash hole interaction) in the neutral molecules. 

The paper is organised as follows. Section~\ref{computational} contains the
technical details for the calculation of the ground\textendash state { properties} 
(Sect.~\ref{dft}) and the electronic absorption spectra (Sect.~\ref{td_dft}).
The results { we} obtained are presented in Sect.~\ref{results} and 
discussed in Sect.~\ref{discussion}. Our concluding remarks are reported in 
Sect.~\ref{conclusion}.

\section{Computational details}
\label{computational}
\subsection{DFT calculations}
\label{dft}
The calculation of the excitation energies and the electronic absorption 
spectra required the previous knowledge of the ground\textendash state optimised 
geometries. For this part of the work we used the Gaussian\textendash based DFT module 
of the \textsc{nwchem} package \cite{apr05}. Geometry optimisations were 
performed using a basis\textendash set with the smallest addition of diffuse functions, 
namely the \mbox{6\textendash31+G$^\star$} basis, a valence double zeta set augmented with 
$d$ polarisation functions and $s$ and $p$ diffuse functions for each carbon 
atom.  

We used the hybrid B3LYP functional, a combination of the Becke's three 
parameter exchange functional \cite{bec93} and the Lee\textendash Yang\textendash Parr 
gradient\textendash corrected correlation functional \cite{lee88}. Although hybrid DFT 
functionals are computationally more expensive than other exchange\textendash correlation
functionals in the local density or generalised gradient 
approximations, B3LYP results for ground\textendash state properties are known to be 
markedly more accurate compared with experiment for a large number of systems 
including PAHs in general \cite{mar96,lan96} and oligoacenes in particular 
\cite{wib97,del01,del03,kad06}. This is confirmed for neutral benzene, whose 
optimised bond\textendash lengths we obtain at the \mbox{B3LYP/6\textendash31+G$^\star$} level are 
r$_\mathrm{CC}$ = 1.399 \AA, and r$_\mathrm{CH}$ = 1.087 \AA, to be compared with 
{ the empirical equilibrium (r$_e$) recommended values of, respectively, 
1.3914$\pm$0.0010 \AA{} and 1.0802$\pm$0.0020 \AA{} \cite{gau00}}. Analogously, the 
{ ground\textendash state} rotational constant we found is 
{ $\sim$5700 MHz}, in agreement with the 
{ empirical equilibrium (B$_e$) determination of 5731.73 MHz \cite{gau00}}.

From the structural relaxations performed for both neutral and charged systems,
we computed via total\textendash energy differences the adiabatic electron affinities and
the adiabatic single and double ionisation energies. At the optimised geometry 
of the neutral molecule we evaluated also the vertical electron affinity 
(EA$_\mathrm{v}$) and the vertical first ionisation energy (IE$_\mathrm{v}$). This
enabled us to obtain the quasiparticle (QP) { corrected HOMO\textendash LUMO gap} 
of the neutral systems considered, which is rigorously defined within the 
{ $\Delta$SCF} scheme \cite{jon89} as:
\begin{equation}
\label{delta}
\mathrm{QP}_\mathrm{gap}^1 = \mathrm{IE}_\mathrm{v} - \mathrm{EA}_\mathrm{v} = 
\mathrm{E}_\mathrm{N+1} + \mathrm{E}_\mathrm{N-1} - 2 \mathrm{E}_\mathrm{N},
\end{equation}
E$_\mathrm{N}$ being the total energy of the { N\textendash electron system}. We used 
also the following approximate expression: 
\cite{god88}:
\begin{equation}
\label{delta2}
\mathrm{QP}_\mathrm{gap}^2 = \epsilon_{N+1}^{N+1} - \epsilon_{N}^{N},
\end{equation}
where $\epsilon_\mathrm{i}^\mathrm{j}$ is the i$^\mathrm{th}$ eigenvalue of the 
j\textendash electron system. { The results obtained using the above 
Eqs.~(\ref{delta}) and (\ref{delta2}) tend to coincide as the system gets 
larger and the orbitals more delocalised}. The \mbox{B3LYP/6\textendash31+G$^\star$} level 
of theory shows good agreement with experiments for the electron affinities of 
PAHs \cite{mal05,des00,mod06}, but it is known to be unable to predict their 
absolute ionisation energies with chemical accuracy ($\pm$0.1eV) 
\cite{mal07a,wib97,kad06,sch01}. This has been discussed in detail by Kadantsev
et al. \cite{kad06}, who concluded that a better description of the electron 
correlation is needed to reproduce the experimental IEs.  { Switching to 
another computational scheme one could employ many\textendash body perturbation theory 
in the so\textendash called Hedin's GW approximation \cite{hed65}. This method, in which 
the QP energies are calculated from the self\textendash energy operator of the system 
(given as the product of the Green's function $\mathcal{G}$ and the screened 
Coulomb interaction $W$), gives results in excellent 
agreement with the available experiments for many materials (see e.g., 
\cite{hib86}). To assess the reliability of our $\Delta$SCF QP-corrected HOMO-LUMO 
gaps, we compared them with the GW results obtained for the oligoacenes 
\cite{nie05}.} The { QP\textendash corrected} HOMO\textendash LUMO gap of our benchmark benzene 
molecule is 10.59~eV (IE$_\mathrm{v}$ = 9.20~eV, EA$_\mathrm{v}$ = -1.39~eV), that
compares favorably with { the GW results \cite{nie05} of 10.59~eV 
(first\textendash principles calculations using Gaussian\textendash type orbitals) and 10.46~eV 
(DFT\textendash based tight\textendash binding calculations), respectively, and with the 
experimental value of $\sim$10.36~eV (IE$_\mathrm{v}$ = 9.24384$\pm$0.00006 
\cite{nem93}, EA$_\mathrm{v}$ = -1.12$\pm$0.03~eV \cite{bur87}).}

\subsection{TD\textendash DFT calculations}
\label{td_dft}
Thanks to the good compromise between accuracy and computational costs, 
compared to many\textendash electron wavefunction\textendash based \emph{ab initio} methods,
TD\textendash DFT is the most widely used approach to compute the excitation energies of 
such complex molecules as PAHs 
\cite{hei00,ngu02,par03,gri03,die04a,die04b,hou01,hir99,hal00,hir03,bau05,
series}. In this study we used two different implementations of TD\textendash DFT in the 
linear response regime, in conjunction with different representations of the 
wavefunctions:
\begin{enumerate}
\item \label{one}
the real\textendash time propagation scheme using a grid in real space \cite{yab99}, as 
implemented in the \textsc{octopus} computer program \cite{octopus}.
\item \label{two}
the frequency\textendash space implementation \cite{hir99b} based on the linear 
combination of localised orbitals, as given in the \textsc{nwchem} 
package \cite{apr05}.
\end{enumerate}
In the first scheme (\ref{one}) the time\textendash dependent Kohn\textendash Sham equations are 
directly solved in real time and the wavefunctions are represented by their 
discretised values on a uniform spatial grid. The static Kohn\textendash Sham 
wavefunctions are perturbed by an impulsive electric field and propagated for a
given finite time interval. In this way, all of the frequencies of the system
are excited. The whole absolute photo\textendash absorption cross\textendash section { $\sigma(E)$} 
then follows from the dynamical polarisability $\alpha(E)$, which is related 
to the Fourier transform of the time\textendash dependent dipole moment of the molecule. 
The relation is:
\begin{equation}
\label{sigma}
\sigma(E) = \frac{8 \pi^2 E}{h c}\,\Im \{\alpha(E)\},
\end{equation}
where $h$ is Planck's constant, { $\Im \{\alpha(E)\}$ is the imaginary part of
the dynamical polarisability}, and $c$ the velocity of light in vacuum. The 
dipole strength\textendash function $S(E)$ is related to $\sigma(E)$ by the equation: 
\begin{equation}
\label{strength}
S(E) = \frac{m_e c} {\pi h e^2} \sigma(E),
\end{equation}
{ $m_e$ and $e$ being respectively the mass and charge of the electron}. 
{ $S(E)$} has units of oscillator strength per unit energy and satisfies 
the Thomas\textendash Reiche\textendash Kuhn dipole sum\textendash rule 
\mbox{$N_\mathrm{e} = \displaystyle \int\!dE\,S(E)$}, where $N_\mathrm{e}$ is 
the total number of electrons. The great advantage of obtaining the whole 
response at once is particularly useful for astrophysical applications, for 
which the whole absorption spectrum is needed.

In the most widely used frequency\textendash space TD\textendash DFT implementation (\ref{two}), 
based on the linear response of the density\textendash matrix, the poles of the 
linear response function correspond to vertical excitation energies and the 
pole strengths to the corresponding oscillator strengths \cite{cas95}. With 
this method computational costs scale steeply with the number of required 
transitions and electronic excitations are thus usually limited to the 
low\textendash energy part of the spectrum. From a computational point of view the 
advantages of the real\textendash time propagation method are discussed, e.g., in 
Ref.~\cite{lop05}. On the other hand, the main drawbacks of the real\textendash time
approach are that: (i) no information is given on dipole\textendash forbidden 
singlet\textendash singlet and singlet\textendash triplet transitions, and (ii) one does not 
obtain independent { information} for each excited state, such as its { 
irreducible} representation of the point group of the given molecular system, 
and the description of the excitations in terms of promotion of electrons in 
an orbital picture.

We performed the \textsc{octopus} calculations in the local\textendash density 
approximation, with the exchange\textendash correlation energy density of 
the homogeneous electron gas \cite{cep80} parametrised by Perdew \& Zunger 
\cite{per81}. The ionic potentials are replaced by norm\textendash conserving 
pseudo\textendash potentials \cite{tro91}. We used a grid spacing of 0.3~\AA{} and 
determined the box size by requiring each atom to be at least 4~\AA{} away from 
its edges. We furthermore added a 1~\AA{} thick absorbing boundary, which quenches 
spurious resonances due to standing waves in the finite simulation box 
used to confine the molecules \cite{yab99,mar03}. We used a time integration 
length T=20 $\hbar/$eV, corresponding to an energy resolution of $\hbar/T$=0.05~eV. 
For the numerical integration of the time evolution we used a time step of 
0.002~$\hbar/$eV, which ensured energy conservation with good numerical accuracy. 

The TD-DFT calculations with \textsc{nwchem} were performed at the same level 
B3LYP/\mbox{6-31+G$^\star$} used to obtain the ground\textendash state geometries. Although 
basis set convergence is not yet expected at the level we used, our results for
the neutral systems are almost coincident with the ones obtained in 
Ref.~\cite{kad06} using the larger \mbox{6\textendash311++G$^{\star\star}$} basis, which is 
supplemented with a third layer of valence functions and includes polarisation 
and diffuse functions on both carbon and hydrogen atoms. We thus believe our 
theoretical predictions to be sufficiently accurate for the purposes of this
work. In the case of neutral benzene, we predict the strong $\pi\to\pi^\star$ transition
$^1$A$_\mathrm{1g}\to^1$E$_\mathrm{1u}$ at 6.96~eV with an oscillator strength of 
1.22, in good agreement with the measured band position in vapour\textendash phase of 
6.94~eV with an $f$\textendash value of 1.2~\cite{pic51}.

In order to assess the { choice} of this specific exchange\textendash correlation 
functional, at the B3LYP optimised geometries, we used also the 
gradient\textendash corrected BLYP functional \cite{bec88,lee88}, the same approach
used in previous studies of PAH ions \cite{hir99,hal03,hir03,bau05}. { With 
both methods we restricted ourselves to the first 20 singlet\textendash singlet roots.} 
Since we are interested in the behaviour of the lowest\textendash lying permitted 
in\textendash plane long\textendash polarised and short\textendash polarised electronic transitions as a 
function of charge state and molecular size, { in the following we report
only the first five electronic transitions.} 
{ The complete set of electronic excitation energies and oscillator 
strengths computed at both B3LYP and BLYP levels, including also the optically 
inactive ones, are available in our online database of the computed spectral 
properties of PAHs \cite{mal07b}.}

\section{Results}
\label{results}
\subsection{Static properties}

\begin{figure}[!]
\includegraphics{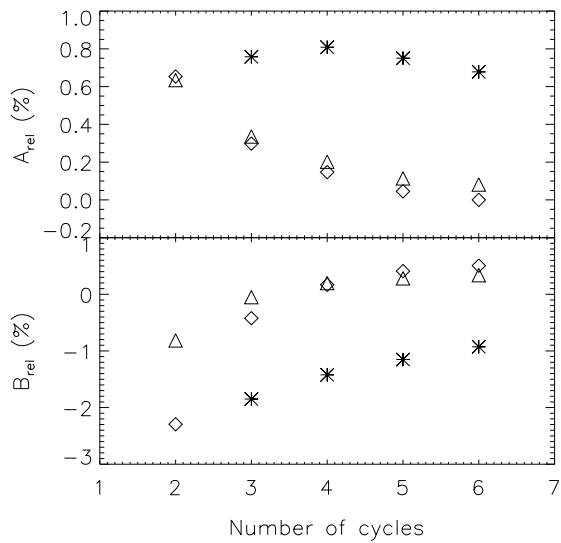}
\caption{{ Percentage} variation of the rotational constants A (top panel) 
and B (bottom panel) relative to the neutral counterparts for anions 
(asterisks), cations (triangles), and dications (diamonds), as a function of 
molecular size. We omit the entries for naphthalene anion, which is known not 
to form a stable anion in the gas\textendash phase.}
\label{rotconst}
\end{figure}

Geometry optimisations { with} \textsc{nwchem} were performed using tight 
convergence criteria, { that are specified by maximum and root mean square 
gradient thresholds of 1.5$\cdot$10$^{-5}$ and 1.0$\cdot$10$^{-5}$ atomic units, 
respectively, and maximum and root mean square thresholds of the Cartesian step
respectively of 6.0$\cdot$10$^{-5}$ and 4.0$\cdot$10$^{-5}$ atomic units. According to 
previous studies \cite{nie95,hir99} the lifting of the molecular symmetry
D$_\mathrm{2h}$ of the neutral molecules is not expected to lead to optimised 
geometries with lower symmetry for the corresponding charged species. We 
indeed confirmed that the structural parameters obtained for naphthalene,
anthracene, and tetracene in the charge\textendash states -1,+1, and +2, ignoring any 
apparent symmetry and adopting the D$_\mathrm{2h}$ symmetry, are coincident 
within numerical errors. We therefore assumed the above D$_\mathrm{2h}$
constraint for all subsequent calculations in the paper.}
Our \mbox{B3LYP/6-31+G$^\star$} geometry optimisations for the neutral molecules 
give structural parameters in good agreement with those previously published 
\cite{wib97,del03,kad06}. In particular, although a larger basis was used in 
Ref.~\cite{kad06}, the two sets of results are almost coincident and compare 
fairly well with the available x\textendash ray data. 
We do not discuss here the changes 
of the single bond lengths and bond angles occurring in the charged species, 
compared to the corresponding neutral ones. { Depictions of the structures
of each molecule considered, in which internal coordinates are shown and
compared, are given as supplementary material to the paper.}
Instead, Fig.~\ref{rotconst} 
presents in a 
collective way the structural variations relative to the neutral molecules, 
expressed in terms of the { percentage} variations of the rotational 
constants A and B. These latter quantities are proportional to the inverse of 
the principal momenta of inertia corresponding to the in\textendash plane short and 
long\textendash axis of the molecules, respectively: A=(h/8$\pi^2$c)I$_\mathrm{short}$, 
B=(h/8$\pi^2$c)I$_\mathrm{long}$. Note that in Fig.~\ref{rotconst} we omitted the
results corresponding to naphthalene anion, { since gas\textendash phase naphthalene 
is unable to bind an additional electron in its LUMO state \cite{bur87}}. 
The ground\textendash state optimised geometries of all of the molecules considered, 
{ both Cartesian and internal coordinates}, are freely available in our 
online database of the computed spectral properties of PAHs \cite{mal07b}.

\begin{figure}
\includegraphics{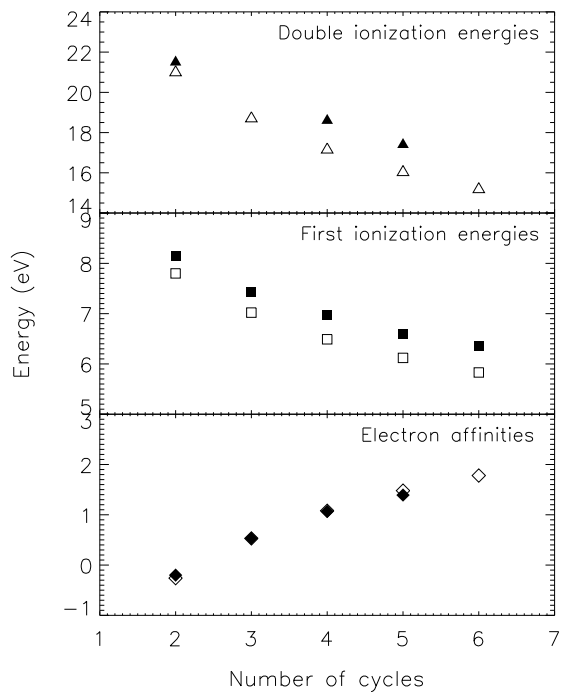}
\caption{{ Computed adiabatic ionisation energies and electron affinities 
of the studied oligoacenes as a function of size. The corresponding 
experimental values are represented by the filled symbols (see 
Table~\ref{table}).}}
\label{ioniz}
\end{figure}

All neutral and singly\textendash charged species were computed as singlet and doublet, 
respectively, while for dications we computed both their singlet and triplet 
ground\textendash states. The adiabatic and vertical values of electron affinities and 
single and double ionisation energies as obtained via total energy differences 
are given in Table~\ref{table} and compared with the available experimental 
data \cite{lia05,tob94}. { As shown in Table~\ref{table}, for all of the 
five molecules considered in their doubly\textendash ionised state, our calculations 
predict the total energy of the singlet state to be lower than that of the 
triplet state by $\sim$0.5\textendash1.0~eV. Figure~\ref{ioniz} displays the computed 
electron affinities and first and double ionisation energies as a function of 
size, and compares them with the available laboratory data.}

\begin{landscape}

\begin{table}
\caption{Adiabatic and vertical values (in parentheses), all data in eV, of 
electron affinities and single and double ionisation energies of the 
oligoacenes considered in this work (C$_{4n+2}$H$_{2n+4}$, n=2, 3, 4, 5, 6)
as obtained through total energy differences at the \mbox{B3LYP/6\textendash31+G$^\star$} 
level. For comparison we list also the experimental { adiabatic} electron 
affinities and { adiabatic} single ionisation energies taken from the NIST 
Chemistry WebBook \cite{lia05}, as well as the { adiabatic} second 
ionisation energies photon\textendash impact measurements \cite{tob94}.}
\begin{center}
\begin{tabular}{c c c c c c c c}
\hline \hline
{ Number} & \multicolumn{2}{c}{{ Electron affinity}} & 
\multicolumn{2}{c}{{ First ionisation energy}} &
{ Dication} & \multicolumn{2}{c}{{ Double ionisation energy}}\\
{ of cycles} & Ad.(Vert.) & Exp. & Ad.(Vert.) & Exp. & { state} &
Ad.(Vert.) & Exp. \\
\hline 
\multirow{2}*{2} & 
\multirow{2}*{-0.26(-0.38)} & 
\multirow{2}*{-0.20$\pm$0.05$^a$} & 
\multirow{2}*{7.80(7.89)} & 
\multirow{2}*{8.144$\pm$0.001$^b$} & { singlet} &
\multirow{1}*{20.99(21.35)} & 
\multirow{2}*{21.5$\pm$0.2} \\
& & & & & { triplet} & \multirow{1}*{21.45(21.57)}&\\
\multirow{2}*{3} & 
\multirow{2}*{0.53(0.43)} & 
\multirow{2}*{0.530$\pm$0.005$^c$} & 
\multirow{2}*{7.02(7.09)} &  
\multirow{2}*{7.439$\pm$ 0.006$^d$} & { singlet} &
\multirow{1}*{18.70(18.95)} & 
\multirow{2}*{\textemdash}\\  
& & & & & { triplet} & \multirow{1}*{19.70(19.80)}&\\
\multirow{2}*{4} & 
\multirow{2}*{1.08(1.00)} & 
\multirow{2}*{1.067$\pm$0.043$^e$} & 
\multirow{2}*{6.49(6.55)} & 
\multirow{2}*{6.97$\pm$0.05$^f$} & { singlet} &
\multirow{1}*{17.15(17.34)} & 
\multirow{2}*{18.6$\pm$ 0.2}\\
& & & & & { triplet} & \multirow{1}*{17.96(18.11)}&\\
\multirow{2}*{5} & 
\multirow{2}*{1.48(1.41)} & 
\multirow{2}*{1.392$\pm$0.043$^e$} & 
\multirow{2}*{6.12(6.16)} & 
\multirow{2}*{{ 6.589$\pm$0.001$^g$}} & { singlet} &
\multirow{1}*{16.03(16.18)} & 
\multirow{2}*{17.4$\pm$0.2}\\
& & & & & { triplet} & \multirow{1}*{16.67(16.80)}&\\
\multirow{2}*{6} & 
\multirow{2}*{1.78(1.72)} & 
\multirow{2}*{\textemdash} & 
\multirow{2}*{5.83(5.87)} & 
\multirow{2}*{6.36$\pm$0.02$^f$} & { singlet} &
\multirow{1}*{15.18(15.30)} & 
\multirow{2}*{\textemdash}\\
& & & & & { triplet} & \multirow{1}*{15.68(15.78)}&\\
\hline
\end{tabular}
\end{center}
\label{table}
\smallskip
{
$^a$ Extrapolated from the EAs of naphthalene\textendash water clusters
determined via photoelectron spectroscopy \cite{sch00}.\\
$^b$ From laser threshold photoelectron spectroscopy \cite{coc93}.\\
$^c$ From photodetachment photoelectron spectroscopy \cite{sch97}.\\
$^d$ From two\textendash laser photoionisation supersonic jet mass 
spectrometry  \cite{hag88}.\\
$^e$ Estimated from gas\textendash phase electron attachment free energies with the
electron\textendash transfer equilibria technique \cite{cro93}.\\
$^f$ From gas\textendash phase photoelectron spectroscopy \cite{sch77}.\\
$^g$ From high\textendash resolution gas\textendash phase photoelectron spectroscopy \cite{gru02}.}
\end{table}

\end{landscape}

\begin{figure}
\includegraphics{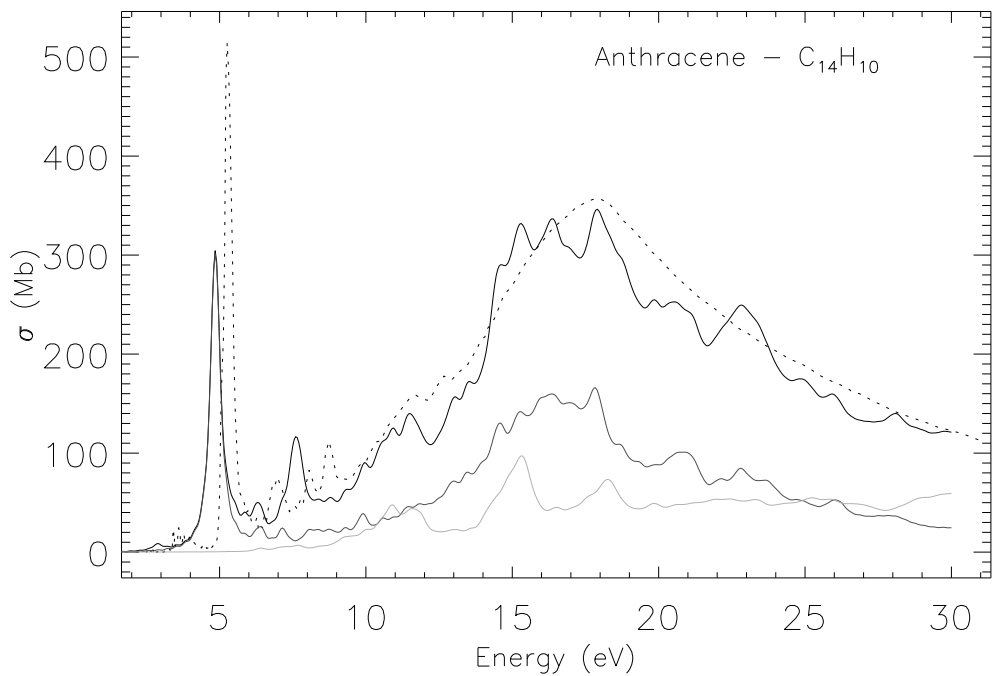}
\caption{Comparison between the computed (solid black line) photo\textendash absorption
cross\textendash section $\sigma(E)$ of neutral anthracene (C$_{14}$H$_{10}$) and the 
corresponding gas\textendash phase absorption spectrum (dotted line, taken from 
Refs.~\cite{job92a,job92b}). 
The contributions corresponding to polarisations along the $x$\textendash axis 
(in\textendash plane short), $y$\textendash axis (in\textendash plane long), and $z$\textendash axis (out\textendash of\textendash plane),
are marked in gray, dark gray, and light gray, respectively. Units are 
megabarns, 1Mb~=~10$^{-18}$~cm$^2$.}
\label{anthr}
\end{figure}

\begin{figure}
\includegraphics{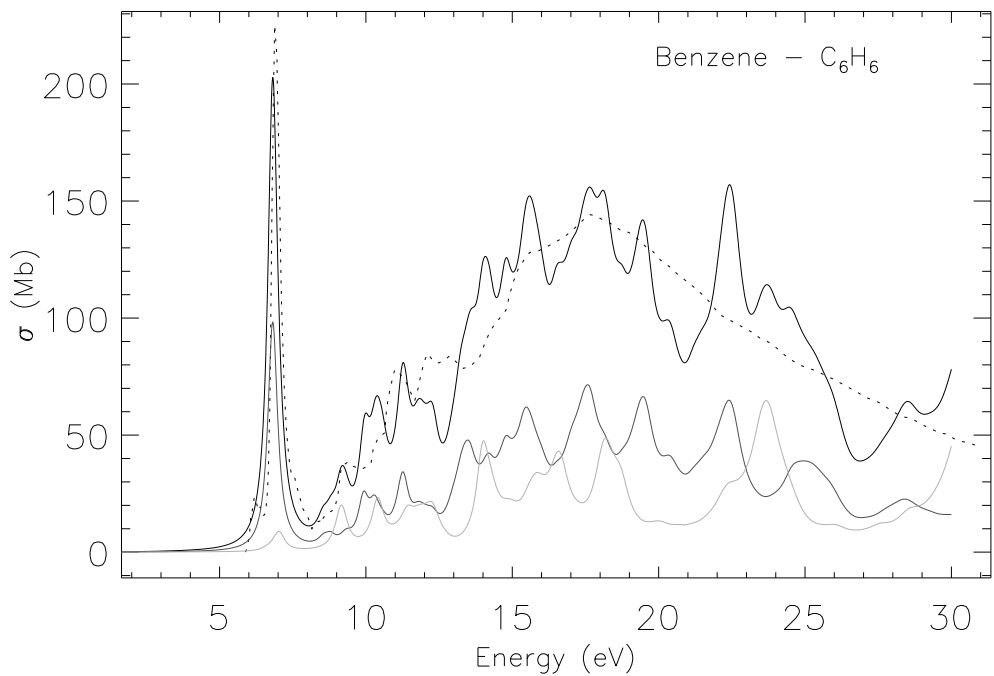}
\caption{Same as Fig.~\ref{anthr} for neutral benzene. Note that, due to the 
symmetry of the molecule, the curves corresponding to $x$ and $y$ 
polarisations are obviously coincident.}
\label{benz}
\end{figure}

\begin{figure}
\includegraphics{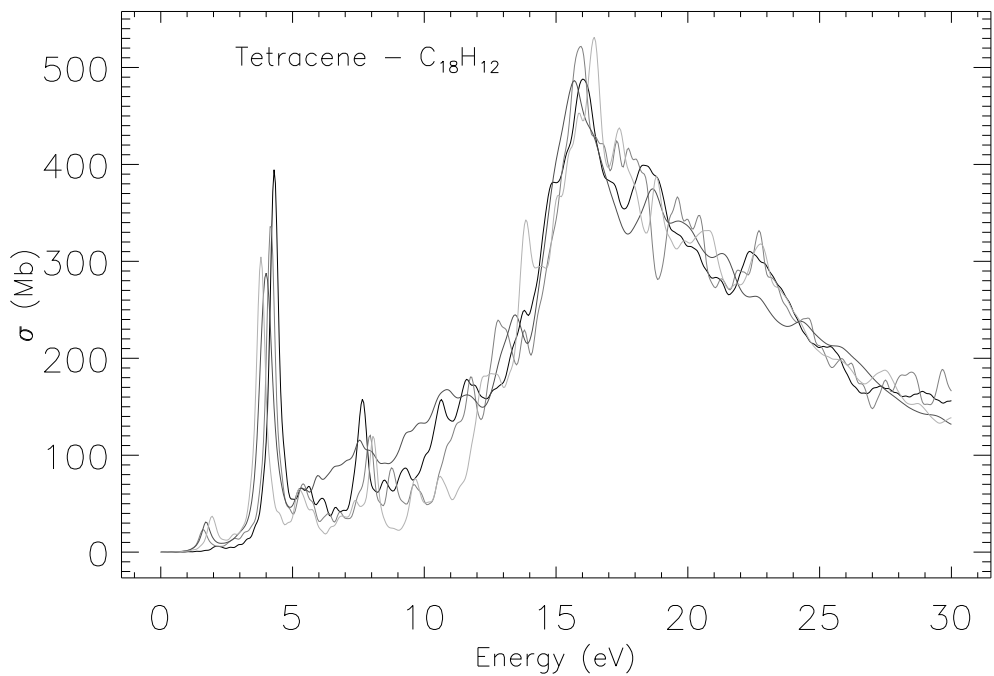}
\caption{Computed photo\textendash absorption cross\textendash section $\sigma(E)$ of tetracene 
(C$_{18}$H$_{12}$) in neutral (black), anionic (dark gray), cationic (gray) and 
dicationic (light gray) charge state, as obtained with the real\textendash time 
real\textendash space implementation of TD-DFT.}
\label{comp1}
\end{figure} 

\begin{figure}
\includegraphics{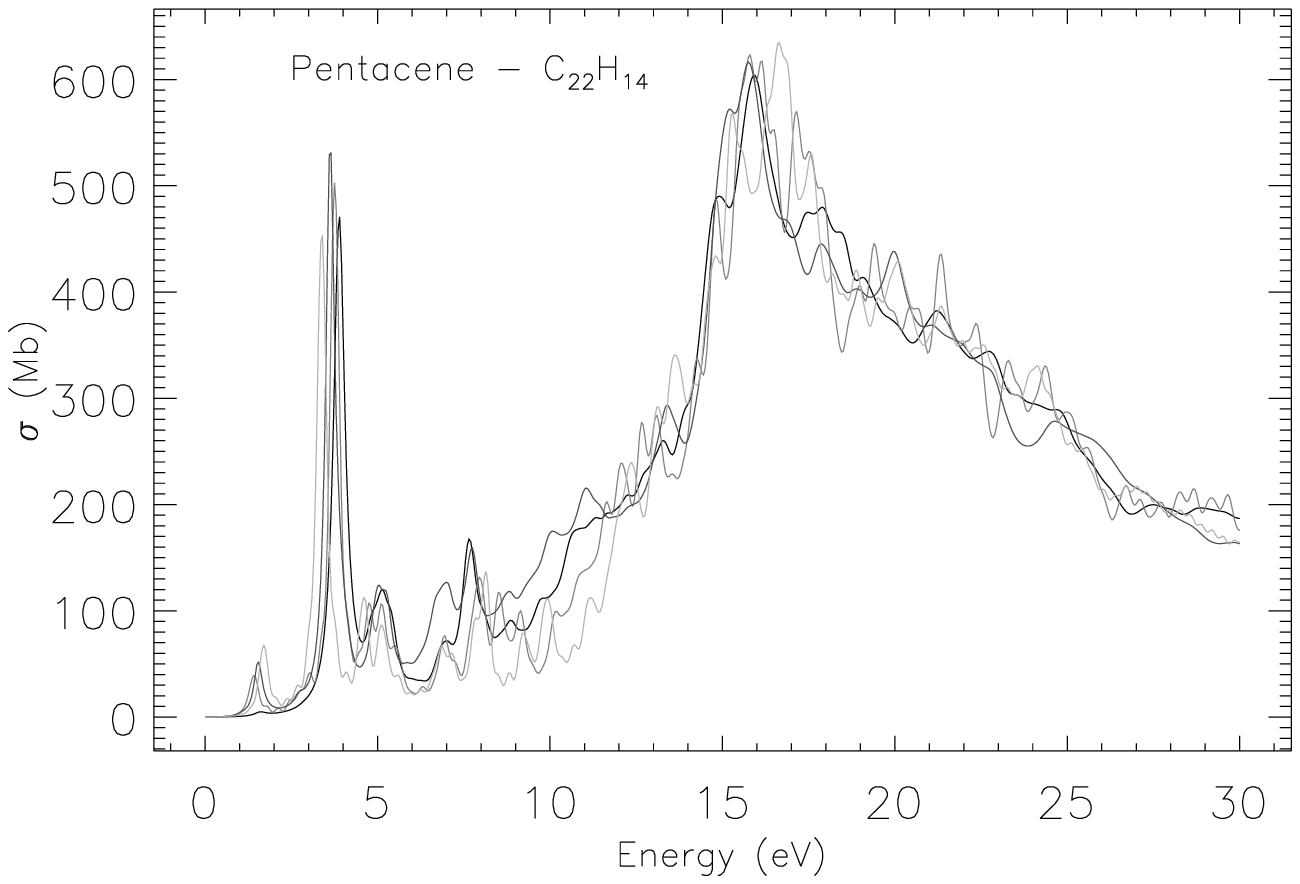}
\caption{Same as Fig.~\ref{comp1} for pentacene (C$_{22}$H$_{14}$).}
\label{comp2}
\end{figure} 

\subsection{Excitation energies and electronic absorption spectra}

\begin{figure}
\includegraphics{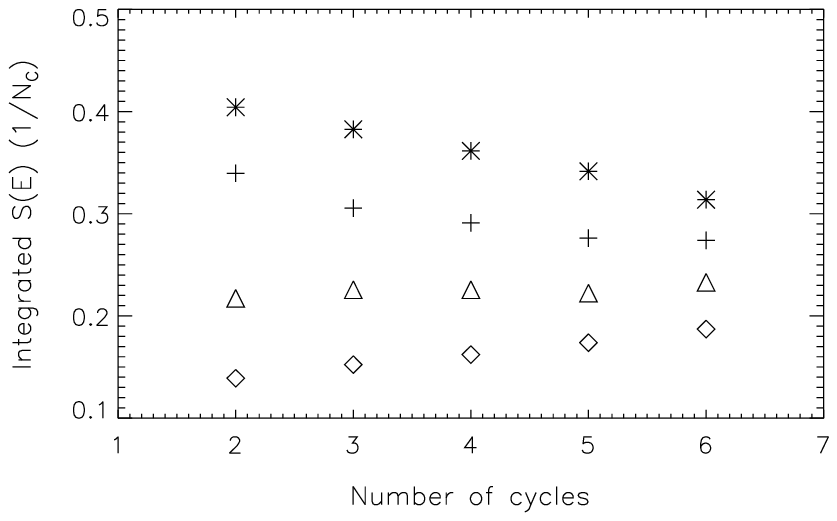}
\caption{Comparison between the integrated values in the range 6\textendash12~eV
of the individual dipole strength\textendash functions $S(E)$ (see Eq.~\ref{strength}) 
divided by the number of { carbon atoms} for the oligoacenes 
anions (asterisks), neutrals (crosses), cations (triangles), and dications 
(diamonds) considered, as a function of molecular size.}
\label{integrated}
\end{figure}

Since its first applications, the real\textendash time TD\textendash DFT method in real space was 
proven to give good results for neutral benzene \cite{yab99,mar03}, compared 
with the experimental spectrum recorded in the energy range 6\textendash35 eV 
\cite{koc72a}. Applying this same approach to a large sample of PAHs, we 
already showed \cite{mal04,mul06b} our results to be in good agreement up to 
photon energies of about $30$~eV with the experimental data obtained for a few 
neutral PAHs with the synchrotron radiation facility of SUPERACO 
\cite{job92a,job92b}. The comparison between the theoretical spectra obtained 
with the \textsc{octopus} code and the experimental photo\textendash absorption 
cross\textendash sections of neutral anthracene 
(C$_{14}$H$_{10}$) and benzene in the gas\textendash phase are shown in Figs.~\ref{anthr} and
\ref{benz}. The latter experimental spectra are in good agreement with the ones
obtained with the synchrotron radiation from the electron accelerator DESY for 
anthracene \cite{koc73}, and benzene \cite{koc72a}, respectively. Since the 
theoretical spectrum is averaged over the three $x$, $y$, and $z$ 
polarisations, each single contribution is also shown in Figs.~\ref{anthr} and
\ref{benz}. As examples, the spectra computed for tetracene (C$_{18}$H$_{12}$) 
and pentacene (C$_{22}$H$_{14}$) in the four charge states considered are given 
in Figs.~\ref{comp1} and \ref{comp2}. The spectra for all of the other 
molecules under study can be found in our online database \cite{mal07b}.
Figure~\ref{integrated} shows the comparison between the integrated values 
in the range 6\textendash12~eV of the individual dipole strength\textendash functions $S(E)$ 
(see Eq.~\ref{strength}) { divided by the number of carbon atoms in each 
molecule} as a function of molecular size. 

The first few permitted electronic transitions of each molecule, as obtained 
at the \mbox{B3LYP/6-31+G$^\star$} and \mbox{BLYP//B3LYP/6-31+G$^\star$} levels with the 
TD\textendash DFT frequency\textendash space implementation of \textsc{nwchem} are reported in 
Tables~\ref{naphthalene}-\ref{hexacene}, and compared with the available 
experimental data that we could find in the literature. We deliberately omitted
the large amount of photoelectron data available for neutral species \cite{pe}.
The use of such data for spectral assignments of the so\textendash called Koopmans 
transitions of radical cations has been already discussed many times 
\cite{bal98,hir99,hal00,hir03}. 
{ Electronic excited states are classified under the point\textendash group 
D$_\mathrm{2h}$ and the ground\textendash state symmetry is specified for each 
charge\textendash state. As to the character of the excited electronic states, we
analyse the nature of the corresponding transitions in terms of the occupied 
and virtual molecular orbitals that have been interchanged between the 
ground and the excited electronic states \cite{dre05}. The above description is
computationally convenient and straightforward for states which are well 
described with only one or two significantly contributing ``excited'' Slater 
determinants \cite{dre05}. This is the case for the electronic excited states 
reported in Tables~\ref{naphthalene}-\ref{hexacene}. The use of more 
sophisticated and physically more appealing ways to obtain information about 
an electronic transition, such as difference density or attachment/detachment 
density plots \cite{dre05}, is outside the scope of the present work.
We used the same notation as in Refs.~\cite{hir99,hir03} where the $\pi$ orbitals
are numbered in the order of increasing energies and $\pi_{-1}$, $\pi_0 (\pi_0^\star)$, and 
$\pi_1^\star$ denote the highest doubly occupied $\pi$ orbital, the singly occupied 
(unoccupied) $\pi$ orbital, and the lowest doubly unoccupied $\pi$ orbital, 
respectively.}

\begin{figure}
\includegraphics{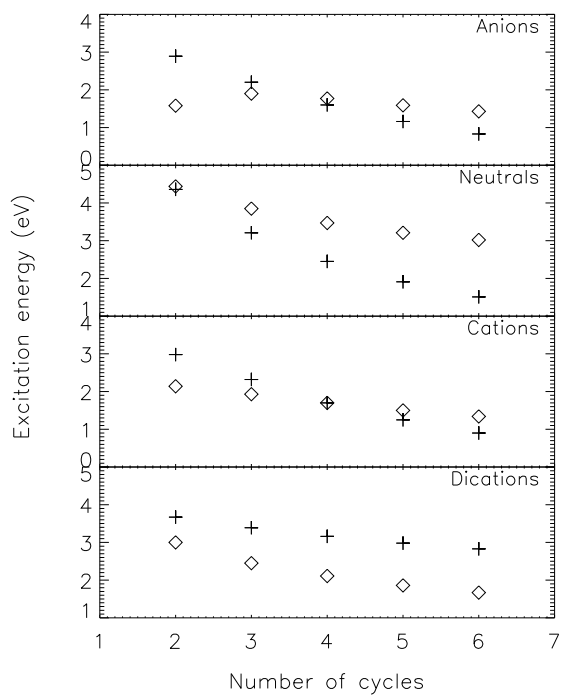}
\caption{Calculated positions (\mbox{B3LYP/6-31+G$^\star$}) of the lowest 
short\textendash polarised (crosses, $p$\textendash bands in the neutral species) and lowest 
long\textendash polarised (diamonds, $\alpha$\textendash bands in the neutrals) electronic transitions in
the five oligoacenes considered as a function of the size and charge\textendash state of 
the molecule.}
\label{trend1}
\end{figure}

\begin{figure}
\includegraphics{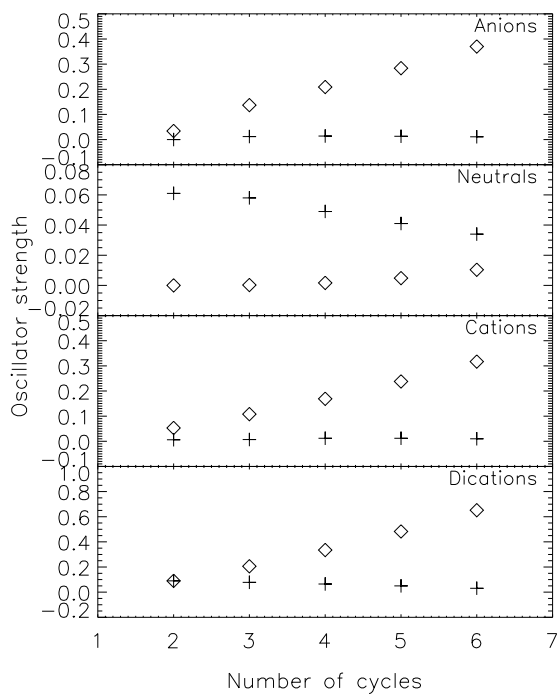}
\caption{Same as Fig.~\ref{trend1} for the corresponding oscillator strengths.}
\label{trend2}
\end{figure}

Figures \ref{trend1} and \ref{trend2} display, respectively, 
the \mbox{B3LYP/6-31+G$^\star$} positions and the corresponding oscillator 
strengths of the lowest in\textendash plane short\textendash polarised (p\textendash bands in the neutral 
molecules) and long\textendash polarised ($\alpha$\textendash bands in the neutral molecules) electronic 
transitions as a function of the number of benzene units and the charge state 
of the molecule.

\begin{figure}
\includegraphics{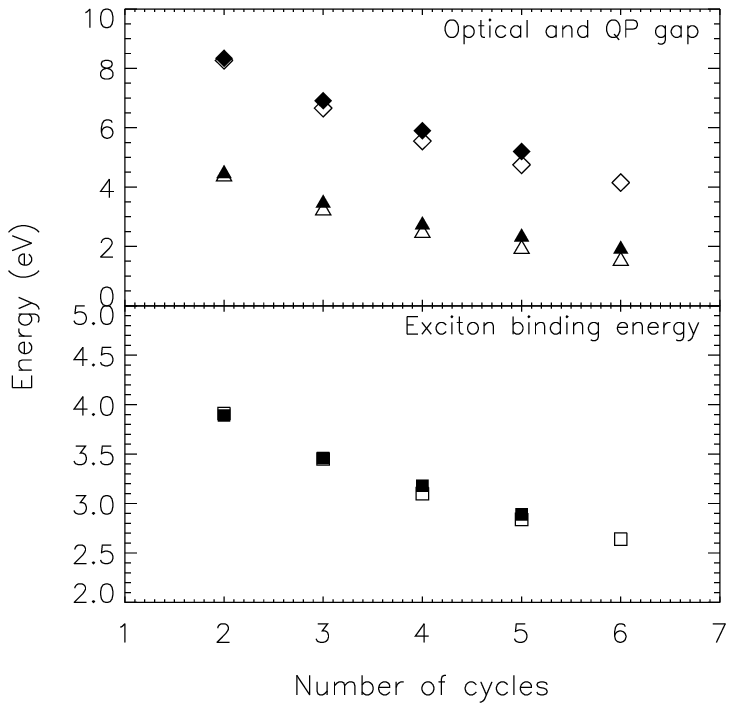}
\caption{Top panel: optical gap E$_\mathrm{gap}^\mathrm{TD-DFT}$ as obtained
via TD\textendash DFT at the \mbox{B3LYP/6-31+G$^\star$} level (open triangles) and { 
$\Delta$SCF QP\textendash corrected HOMO\textendash LUMO gap QP$^1_\mathrm{gap}$ computed via 
Eq.~(\ref{delta})} (open diamonds), as a function of molecular size.  
Bottom panel: exciton binding energy E$_\mathrm{bind}$ estimated as 
QP$^1_\mathrm{gap}$ - E$_\mathrm{gap}^\mathrm{TD-DFT}$ (open squares). In both 
figures, the corresponding experimental values are represented by the filled 
symbols (see Table~\ref{table1}).}
\label{gap}
\end{figure}

\begin{table}
\caption{Singlet\textendash singlet lowest\textendash lying permitted excitation energies {\bf
(in eV)} of naphthalene anion, neutral, cation, and dication as obtained via 
frequency\textendash space TD\textendash DFT. {\bf The corresponding oscillator strengths are
given in parentheses.} Polarization of the bands are denoted according to 
Fig.~\ref{geoms} as $x$, $y$, and $z$ for in\textendash plane short, in\textendash plane long, and 
out\textendash of\textendash plane polarized bands, respectively.  {\bf The description of 
each excitation is based on the B3LYP results and is given in terms of the 
occupied and virtual molecular orbitals contributing significantly to it 
\cite{hir99,hir03}. We report also the available experimental data for 
comparison.}}
\begin{center}
\begin{tabular}{c c c c c}
\hline \hline
State (pol.) & {\bf Excitation} & BLYP & B3LYP & Exp.\\
\hline
\multicolumn{5}{c}{Anion (ground\textendash state $^2$B$_{3g}$)}\\[-1.0mm]
1 $^2$B$_{2u}$ ($z$) &$\pi_0\to\pi_2^\star$ & 0.66(0.001) & 0.99(0.001) &  \textemdash\\[-1.0mm]
1 $^2$B$_{1u}$ ($y$) &$\pi_0\to\pi_7^\star$ & 1.48(0.022) & 1.58(0.034) &$\sim$1.5$^a$\\[-1.0mm]
2 $^2$B$_{2u}$ ($z$) &$\pi_0\to\pi_8^\star$ & 1.85(0.005) & 2.15(0.006) &  \textemdash\\[-1.0mm]
2 $^2$B$_{1u}$ ($y$) &$\pi_0\to\sigma_2^\star$ & 2.30(0.071) & 2.49(0.069) &$\sim$2.5$^a$\\[-1.0mm]
1 $^2$A$_{u}$ ($x$) &$\pi_{-1}\to\pi_1^\star$ & 2.73(0.001) & 2.89($<$0.001) & \textemdash\\
\hline
\multicolumn{5}{c}{Neutral (ground\textendash state $^1$A$_g$)}\\[-1.0mm]
1 $^1$B$_{3u}$ ($x, p$) &$\pi_{-1}\to\pi_1^\star$& 4.06(0.045) & 4.36(0.061) & 
4.45(0.102)$^b$, 4.45(0.109)$^c$, 4.44$^d$\\[-1.0mm]
1 $^1$B$_{2u}$ ($y, \alpha$) &$\pi_{-2}\to\pi_1^\star$, $\pi_{-1}\to\pi_2^\star$ & 4.20($<$0.001) & 
4.44($<$0.001)& 3.97(0.002)$^b$, 4.0$^c$, 3.98$^d$\\[-1.0mm]
2 $^1$B$_{2u}$ ($y, \beta$) &$\pi_{-2}\to\pi_1^\star$, $\pi_{-1}\to\pi_2^\star$ & 5.62(1.155) & 5.85(1.260) & 
5.89(1.3)$^b$, 5.89(1.3)$^c$, 5.86$^d$\\[-1.0mm]
2 $^1$B$_{3u}$ ($x$) &$\pi_{-2}\to\pi_2^\star$ & 5.80(0.146) & 6.08(0.199) & 
6.14(0.3)$^b$, 6.0$^c$, 6.09$^d$\\[-1.0mm]
1 $^1$B$_{1u}$ ($z$) &$\pi_{-2}\to\pi_3^\star$ & 5.77(0.008) & 6.24(0.016) & \textemdash\\
\hline
\multicolumn{5}{c}{Cation (ground\textendash state $^2$A$_u$)}\\[-1.0mm]
1 $^2$B$_{2g}$ ($y$) &$\pi_{-2}\to\pi_0^\star$ & 2.15(0.042) & 2.14(0.053) & 
1.84$^d$, 1.85(0.052)$^e$, 1.84(0.011)$^f$\\[-1.0mm]
1 $^2$B$_{3g}$ ($x$) &$\pi_{-3}\to\pi_0^\star$ & 2.77(0.006) & 2.98(0.006) & 
2.72$^d$, 2.72(0.010)$^e$, 2.69(0.001)$^f$\\[-1.0mm]
2 $^2$B$_{3g}$ ($x$) &$\pi_0\to\pi_1^\star$ & 3.51(0.051) & 
3.59(0.064) & 3.29$^d$, 3.25(0.016)$^f$\\[-1.0mm]
2 $^2$B$_{2g}$ ($y$) &$\pi_{-1}\to\pi_1^\star$ & 3.74(0.012) & 3.91(0.032) & 
4.02$^d$, 4.03(0.024)$^f$\\[-1.0mm]
3 $^2$B$_{2g}$ ($y$) &$\pi_0\to\pi_2^\star$ & 4.30(0.008) & 4.61(0.006) & 
4.49$^d$, 4.52(0.060)$^f$\\
\hline
\multicolumn{5}{c}{Dication (ground\textendash state $^1$A$_g$)}\\[-1.0mm]
1 $^1$B$_{2u}$ ($y$) &$\pi_{-2}\to\pi_1^\star$ & 2.85(0.061) & 3.00(0.090) 
& \textemdash\\[-1.0mm]
1 $^1$B$_{1u}$ ($z$) &$\pi_{-4}\to\pi_1^\star$ & 2.71($<$0.001)& 3.25($<$0.001) & \textemdash\\[-1.0mm]
1 $^1$B$_{3u}$ ($x$) &$\pi_{-3}\to\pi_1^\star$ & 3.32(0.062) & 3.67(0.090) & 
\textemdash\\[-1.0mm]
2 $^1$B$_{2u}$ ($y$) &$\pi_{-1}\to\pi_2^\star$ & 5.06(0.833) & 5.28(0.865) & 
\textemdash\\[-1.0mm]
2 $^1$B$_{1u}$ ($z$) &$\sigma_{-3}\to\pi_1^\star$ & 4.95(0.001) & 5.58(0.002) & \textemdash\\
\hline
\end{tabular}
\end{center}
$^a$Absorption in glassy organic solid \cite{shi73}.
$^b$Gas\textendash phase absorption \cite{geo68}.
$^c$Electron\textendash energy\textendash loss spectroscopy \cite{hub72}.
$^d$Absorption in neon\textendash matrix \cite{sal91}.
$^e$Gas\textendash phase absorption \cite{pin99}.
$^f$Absorption in argon\textendash matrix \cite{kel82}.
\label{naphthalene}
\end{table}

\begin{table}
\caption{Same as Table~\ref{naphthalene} for anthracene in its -1,0,+1, and 
+2 charge states.}
\begin{center}
\begin{tabular}{c c c c c}
\hline \hline
State (pol.) & {\bf Excitation} & BLYP & B3LYP & Exp.\\
\hline
\multicolumn{5}{c}{Anion (ground\textendash state $^2$B$_{1u}$)}\\[-1mm]
1 $^2$A$_g$ ($z$) & $\pi_0\to\pi_1^\star$ & 1.23(0.001) & 1.56(0.001) & \textemdash\\[-1mm]
1 $^2$B$_{3g}$ ($y$) & $\pi_0\to\pi_5^\star$ & 1.83(0.110) & 1.90(0.137) & $\sim$1.7$^a$ \\[-1mm]
2 $^2$A$_g$ ($z$) & $\pi_0\to\pi_7^\star$ & 1.88(0.001) & 2.17(0.002) & \textemdash\\[-1mm]
1 $^2$B$_{2g}$ ($x$) & $\pi{-1}\to\pi_0^\star$& 2.12(0.008)& 2.20(0.012) & $\sim$2.4$^a$\\[-1mm]
2 $^2$B$_{2g}$ ($x$) & $\pi_0\to\pi_{11}^\star$ & 2.78(0.008) & 3.01(0.010) & \textemdash{} \\
\hline
\multicolumn{5}{c}{Neutral (ground\textendash state $^1$A$_g$)}\\[-1mm]
1 $^1$B$_{3u}$ ($x$, $p$) & $\pi_{-1}\to\pi_1^\star$ & 2.92(0.039) & 
3.21(0.058) & 3.27(0.1)$^b$, 3.45$^c$, 3.43$^d$\\[-1mm]
1 $^1$B$_{2u}$ ($y, \alpha$) & $\pi_{-2}\to\pi_1^\star$, $\pi_{-1}\to\pi_2^\star$ & 3.60($<$0.001)& 
3.85($<$0.001) & 3.47-3.60-3.45$^e$, 3.84$^f$\\[-1mm]
2 $^1$B$_{3u}$ ($x$) & $\pi_{-1}\to\pi_6^\star$ & 4.59($<$0.001) & 5.06($<$0.001) & \textemdash\\[-1mm]
2 $^1$B$_{2u}$ ($y, \beta$) & $\pi_{-2}\to\pi_1^\star$, $\pi_{-1}\to\pi_2^\star$& 4.88(1.782) & 5.14(1.992) & 
4.84(2.28)$^b$, 5.24$^c$\\[-1mm]
1 $^1$B$_{1u}$ ($z$) & $\pi_{-1}\to\pi_7^\star$ & 4.83(0.001) & 5.24(0.002) & \textemdash\\
\hline
\multicolumn{5}{c}{Cation (ground\textendash state $^2$B$_{2g}$)}\\[-1mm]
1 $^2$A$_{u}$ ($y$) & $\pi_{-2}\to\pi_0^\star$ & 1.87(0.085) & 1.93(0.108) & 1.71$^a$, 
1.73(0.076)$^g$, 1.75$^h$ \\[-1mm]
1 $^2$B$_{1u}$ ($x$) &  $\pi_0\to\pi_1^\star$ & 2.21(0.004) & 
2.32(0.007) & 2.02(0.018)$^g$\\[-1mm]
2 $^2$B$_{1u}$ ($x$) & $\pi_{-4}\to\pi_0^\star$ & 2.91(0.044) & 
3.17(0.058) & 2.83$^a$, 2.90(0.026)$^g$\\[-1mm]
2 $^2$A$_{u}$ ($y$) & $\pi_{-1}\to\pi_1^\star$ & 3.22(0.017) & 3.39(0.054) & 3.52(0.104)$^g$\\
3 $^2$A$_{u}$ ($y$) & $\pi_0\to\pi_2^\star$& 3.69(0.016) & 4.02(0.010) & 
3.95(0.187)$^g$\\
\hline
\multicolumn{5}{c}{Dication (ground\textendash state $^1$A$_g$)}\\[-1mm]
1 $^1$B$_{2u}$ ($y$) & $\pi_{-2}\to\pi_1^\star$ & 2.31(0.139) & 2.45(0.206) & \textemdash{}
\\[-1mm]
1 $^1$B$_{3u}$ ($x$) & $\pi_{-3}\to\pi_1^\star$ & 3.02(0.052) & 3.39(0.078) & \textemdash\\[-1mm]
2 $^1$B$_{2u}$ ($y$) & $\pi_{-1}\to\pi_2^\star$ & 4.38(1.368) & 4.63(1.444) & \textemdash{}
\\[-1mm]
1 $^1$B$_{1u}$ ($z$) & $\pi_{-8}\to\pi_1^\star$ & 3.92($<$0.001) & 4.66($<$0.001) & \textemdash\\[-1mm]
2 $^1$B$_{3u}$ ($x$) & $\pi_{-4}\to\pi_2^\star$ & 4.96(0.002) & 5.59(0.001) & \textemdash\\
\hline
\end{tabular}
\end{center}
$^a$Absorption in glassy organic solid \cite{shi73}. $^b$Absorption in 
n\textendash heptane solution \cite{kle49}. $^c$Gas\textendash phase absorption \cite{koc73}.
$^d$Fluorescence jet spectroscopy \cite{lam84}.
$^e$Magnetic circular dichroism measurements in solvents 
\cite{mas73,sac73,ste78}. $^f$Two\textendash photon absorption in solution \cite{ber72}. 
$^g$Absorption in argon\textendash matrix \cite{szc93b}. $^h$Gas\textendash phase absorption 
\cite{suk04}.
\label{anthracene}
\end{table}

\begin{table}
\caption{Same as Table~\ref{naphthalene} for tetracene in its -1,0,+1, and 
+2 charge states.}
\begin{center}
\begin{tabular}{c c c c c}
\hline \hline
State (pol.) & {\bf Excitation} & BLYP & B3LYP & Exp.\\
\hline
\multicolumn{5}{c}{Anion (ground\textendash state $^2$B$_{3g}$)}\\[-1mm]
1 $^2$A$_{u}$ ($x$) & $\pi_{-1}\to\pi_0^\star$ & 1.53(0.009) & 1.60(0.014) & 1.69$^a$ \\[-1mm]
1 $^2$B$_{1u}$ ($y$) & $\pi_0\to\pi_1^\star$ & 1.68(0.164) & 1.77(0.209) & 1.50$^a$ \\[-1mm]
1 $^2$B$_{2u}$ ($z$) & $\pi_0\to\pi_3^\star$ & 1.68(0.001) &2.06(0.001) & 1.91$^a$\\[-1mm]
2 $^2$B$_{1u}$ ($y$) & $\pi_0\to\pi_{10}^\star$ & 2.40($<$0.01) & 2.62($<$0.001) & \textemdash{} \\[-1mm]
2 $^2$B$_{2u}$ ($z$) & $\pi_0\to\pi_{11}^\star$ & 2.58(0.001) & 2.91(0.001) & \textemdash\\
\hline
\multicolumn{5}{c}{Neutral (ground\textendash state $^1$A$_g$)}\\[-1mm]
1 $^1$B$_{3u}$ ($x, p$) & $\pi_{-1}\to\pi_1^\star$ & 2.16(0.031) & 2.45(0.049) & 2.62(0.08)$^b$, 
2.60(0.11)$^c$, 
2.72$^d$\\[-1mm]
1 $^1$B$_{2u}$ ($y, \alpha$) & $\pi_{-3}\to\pi_1^\star$, $\pi_{-1}\to\pi_3^\star$ & 3.21(0.001) & 3.47(0.002) & 
3.12$^d$\\[-1mm]
2 $^1$B$_{3u}$ ($x$) & $\pi_{-1}\to\pi_5^\star$ & 3.98($<$0.001) & 4.59($<$0.001) & \textemdash\\[-1mm]
2 $^1$B$_{2u}$ ($y, \beta$) & $\pi_{-3}\to\pi_1^\star$, $\pi_{-1}\to\pi_3^\star$ & 4.34(2.317) & 4.62(2.691) & 
4.55(1.85)$^b$, 4.50(1.75)$^c$
\\[-1mm]
3 $^1$B$_{3u}$ ($x$) & $\pi_{-4}\to\pi_1^\star$ & 4.35(0.010) & 4.92($<$0.001) & 
\textemdash\\
\hline
\multicolumn{5}{c}{Cation (ground\textendash state $^2$A$_{u}$)}\\[-1mm]
1 $^2$B$_{2g}$ ($y$) & $\pi_{-1}\to\pi_0^\star$ & 1.59(0.129) & 1.70(0.169) & 
1.43$^a$, 1.43$^e$\\[-1mm]
1 $^2$B$_{3g}$ ($x$) & $\pi_0\to\pi_1^\star$ & 1.58(0.008) & 1.70(0.012) & 1.65$^a$, 1.66$^e$
\\[-1mm]
2 $^2$B$_{3g}$ ($x$) & $\pi{-4}\to\pi_0^\star$ & 2.67(0.029) & 3.03(0.041) & 
3.14$^a$, 3.16$^e$\\[-1mm]
2 $^2$B$_{2g}$ ($y$) & $\pi{-2}\to\pi_1^\star$ & 2.81(0.025) & 3.05(0.084) & \textemdash\\[-1mm]
3 $^2$B$_{2g}$ ($y$) & $\pi{-5}\to\pi_0^\star$ & 3.17(0.005) & 3.52(0.005) & \textemdash\\ 
\hline
\multicolumn{5}{c}{Dication (ground\textendash state $^1$A$_g$)}\\[-1mm]
1 $^1$B$_{2u}$ ($y$) & $\pi_{-1}\to\pi_1^\star$ & 1.97(0.226) & 2.11(0.335) & \textemdash\\[-1mm]
1 $^1$B$_{3u}$ ($x$) & $\pi_{-4}\to\pi_1^\star$ & 2.80(0.042) & 3.16(0.065) & \textemdash\\[-1mm]
2 $^1$B$_{2u}$ ($y$) & $\pi_{-5}\to\pi_1^\star$ & 3.65(0.093) & 4.05(0.626) & \textemdash\\[-1mm]
1 $^1$B$_{1u}$ ($z$) & $\pi_{-8}\to\pi_1^\star$ & 3.46($<$0.001) & 4.11($<$0.001) & \textemdash\\[-1mm]
3 $^1$B$_{2u}$ ($y$) & $\pi_{-2}\to\pi_2^\star$ & 3.86(1.709) & 4.16(1.353) & \textemdash\\
\hline
\end{tabular}
\end{center}
$^a$Absorption in glassy organic solid \cite{shi73}.
$^b$Absorption in n\textendash heptane solution \cite{kle49}.
$^c$ Absorption in benzene solution \cite{bre60}.
$^d$Absorption in gas\textendash phase \cite{ber71}.
$^e$Absorption in argon\textendash matrix \cite{and85,szc95}.
\label{tetracene}
\end{table}

\begin{table}
\caption{Same as Table~\ref{naphthalene} for pentacene in its -1,0,+1, and 
+2 charge states.}
\begin{center}
\begin{tabular}{c c c c c}
\hline \hline
State (pol.) & {\bf Excitation} & BLYP & B3LYP & Exp.\\
\hline
\multicolumn{5}{c}{Anion (ground\textendash state $^2$B$_{1u}$)}\\[-1mm]
1 $^2$B$_{2g}$ ($x$) & $\pi_{-1}\to\pi_0^\star$ & 1.10(0.008) & 1.16(0.013) & 
1.06$^a$, 1.37$^c$\\[-1mm]
1 $^2$B$_{3g}$ ($y$) & $\pi_0\to\pi_1^\star$ & 1.50(0.224) & 1.60(0.286) & 1.40$^a$, 
1.42$^c$\\[-1mm]
1 $^2$A$_{g}$ ($z$) & $\pi_0\to\pi_3^\star$ & 1.96(0.001) & 2.35(0.001) & \textemdash\\[-1mm]
2 $^2$A$_{g}$ ($z$) & $\pi_0\to\pi_8^\star$ & 2.53($<$0.001) & 2.81(0.001) & \textemdash\\[-1mm]
2 $^2$B$_{3g}$ ($y$) & $\pi{-1}\to\pi_4^\star$ & 2.65(0.016) & 2.86(0.118) & 2.82$^a$\\
\hline
\multicolumn{5}{c}{Neutral (ground\textendash state $^1$A$_g$)}\\[-1mm]
1 $^1$B$_{3u}$ ($x, p$) & $\pi_{-1}\to\pi_1^\star$ & 1.63(0.023) & 
1.91(0.041) & 2.12(0.08)$^b$, 2.28$^c$, 2.31$^d$\\[-1mm]
1 $^1$B$_{2u}$ ($y, \alpha$) & $\pi_{-3}\to\pi_1^\star$, $\pi_{-1}\to\pi_3^\star$ & 2.94(0.004) & 
3.21(0.005) & 3.73$^c$\\[-1mm]
2 $^1$B$_{3u}$ ($x$) & $\pi_{-4}\to\pi_1^\star$ & 3.33($<$0.001) & 3.95($<$0.001) & \textemdash\\[-1mm]
2 $^1$B$_{2u}$ ($y, \beta$) & $\pi_{-3}\to\pi_1^\star$, $\pi_{-1}\to\pi_3^\star$ & 3.93(2.706) & 
4.24(3.346) & 4.00(2.20)$^b$, 4.40$^c$\\[-1mm]
3 $^1$B$_{3u}$ ($x$) & $\pi_{-1}\to\pi_4^\star$ & 3.66(0.021) & 4.26(0.003) & \textemdash\\
\hline
\multicolumn{5}{c}{Cation (ground\textendash state $^2$B$_{2g}$)}\\[-1mm]
1 $^2$B$_{1u}$ ($x$) & $\pi_0\to\pi_1^\star$ & 1.17(0.007) & 1.25(0.012) & 1.27$^c$\\[-1mm]
1 $^2$A$_{u}$ ($y$) & $\pi_{-1}\to\pi_0^\star$ & 1.42(0.182) & 1.50(0.238) & 1.31$^c$\\[-1mm]
2 $^2$A$_{u}$ ($y$) & $\pi_{-2}\to\pi_1^\star$ & 2.61(0.034) & 2.80(0.122) & 2.92$^c$\\[-1mm]
2 $^2$B$_{1u}$ ($x$) & $\pi_{-4}\to\pi_0^\star$ & 2.60(0.019) & 2.85(0.013) & \textemdash\\[-1mm]
2 $^2$B$_{1u}$ ($x$) & $\pi_{-4}\to\pi_0^\star$ & 2.85(0.001) & 3.10(0.022) & \textemdash\\
\hline
\multicolumn{5}{c}{Dication (ground\textendash state $^1$A$_g$)}\\[-1mm]
1 $^1$B$_{2u}$ ($y$) & $\pi_{-1}\to\pi_1^\star$ & 1.73(0.324) & 1.86(0.483) & \textemdash\\[-1mm]
1 $^1$B$_{3u}$ ($x$) & $\pi_{-4}\to\pi_1^\star$ & 2.61(0.029) & 2.98(0.050) & \textemdash\\[-1mm]
2 $^1$B$_{2u}$ ($y$) & $\pi_{-5}\to\pi_1^\star$ & 3.17(0.080) & 3.57(0.240) & \textemdash\\[-1mm]
2 $^1$B$_{3u}$ ($x$) & $\pi_{-3}\to\pi_2^\star$ & 3.20(0.011) & 3.66(0.001) & \textemdash\\[-1mm]
3 $^1$B$_{2u}$ ($y$) & $\pi_{-2}\to\pi_2^\star$ & 3.44(2.022) & 3.75(2.186) & \textemdash\\
\hline
\end{tabular}
\end{center}
$^a$Absorption in glassy organic solid \cite{shi73}.
$^b$Absorption in n\textendash heptane solution \cite{kle49}.
$^c$Absorption in neon\textendash matrix \cite{hal00}.
$^d$Absorption in gas\textendash phase \cite{hei98}.
\label{pentacene}
\end{table}

\begin{table}
\caption{Same as Table~\ref{naphthalene} for hexacene in its -1,0,+1, and 
+2 charge states.}
\begin{center}
\begin{tabular}{c c c c c}
\hline \hline
State (pol.) & {\bf Excitation} & BLYP & B3LYP & Exp.\\
\hline
\multicolumn{5}{c}{Anion (ground\textendash state $^2$B$_{3g}$)}\\[-1mm]
1 $^2$A$_{u}$ ($x$) & $\pi_{-1}\to\pi_0^\star$ & 0.78(0.006) & 0.83(0.011) & \textemdash\\[-1mm]
1 $^2$B$_{1u}$ ($y$) & $\pi_0\to\pi_1^\star$ & 1.34(0.284) & 1.43(0.370) & \textemdash\\[-1mm]
2 $^2$A$_{u}$ ($x$) & $\pi_{-2}\to\pi_1^\star$ & 2.41($<$0.001) & 2.47($<$0.001) & \textemdash\\[-1mm]
1 $^2$B$_{2u}$ ($z$) & $\pi_0\to\pi_3^\star$ & 2.14($<$0.001) & 2.56(0.001) & \textemdash\\[-1mm]
2 $^2$B$_{1u}$ ($y$) & $\pi_{-1}\to\pi_2^\star$ & 2.51(0.043) & 2.71(0.149) & \textemdash\\
\hline
\multicolumn{5}{c}{Neutral (ground\textendash state $^1$A$_{g}$)}\\[-1mm]
1 $^1$B$_{3u}$ ($x, p$) & $\pi_{-1}\to\pi_1^\star$ & 1.24(0.017) & 1.51(0.034) & 1.90$^a$\\[-1mm]
1 $^1$B$_{2u}$ ($y, \alpha$) & $\pi_{-3}\to\pi_1^\star$, $\pi_{-1}\to\pi_3^\star$ & 2.75(0.009) & 
3.02(0.010) & 2.80$^a$\\[-1mm]
2 $^1$B$_{3u}$ ($x$) & $\pi_{-4}\to\pi_1^\star$ & 2.78(0.001) & 3.39($<$0.001) & \textemdash\\[-1mm]
3 $^1$B$_{3u}$ ($x$) & $\pi_{-1}\to\pi_4^\star$ & 3.07(0.028) & 3.69(0.010) & \textemdash\\[-1mm]
4 $^1$B$_{3u}$ ($x$) & $\pi{-2}\to\pi_2^\star$ & 3.35(0.017) & 3.82(0.061) & \textemdash\\
\hline
\multicolumn{5}{c}{Cation (ground\textendash state $^2$A$_{u}$)}\\[-1mm]
1 $^2$B$_{3g}$ ($x$) & $\pi_0\to\pi_1^\star$ & 0.84(0.006) & 0.90(0.010) & \textemdash\\[-1mm]
1 $^2$B$_{2g}$ ($y$) & $\pi_{-1}\to\pi_0^\star$ & 1.27(0.246) & 1.34(0.317) & \textemdash\\[-1mm]
2 $^2$B$_{3g}$ ($x$) & $\pi_{-3}\to\pi_1^\star$ & 2.30($<$0.001) & 2.40($<$0.001) &\textemdash\\[-1mm]
2 $^2$B$_{2g}$ ($y$) & $\pi_{-2}\to\pi_1^\star$ & 2.43(0.046) & 2.63(0.167) & \textemdash\\[-1mm]
3 $^2$B$_{3g}$ ($x$) & $\pi_{-4}\to\pi_0^\star$ & 2.52(0.007) & 2.92(0.021) & \textemdash\\
\hline
\multicolumn{5}{c}{Dication (ground\textendash state $^1$A$_{g}$)}\\[-1mm]
1 $^1$B$_{2u}$ ($y$) & $\pi_{-1}\to\pi_1^\star$ & 1.55(0.440) & 1.67(0.652) & \textemdash\\[-1mm]
1 $^1$B$_{3u}$ ($x$) & $\pi_{-4}\to\pi_1^\star$& 2.43(0.009) & 2.83(0.031) & \textemdash\\[-1mm]
2 $^1$B$_{3u}$ ($x$) & $\pi_{-3}\to\pi_2^\star$ & 2.66(0.018) & 3.05(0.001) & \textemdash\\[-1mm]
2 $^1$B$_{2u}$ ($y$) & $\pi_{-5}\to\pi_1^\star$& 2.82(0.062) & 3.21(0.130) & \textemdash\\[-1mm]
3 $^1$B$_{2u}$ ($y$) & $\pi_{-2}\to\pi_2^\star$ & 3.10(2.203) & 3.44(2.642) & \textemdash\\
\hline
\end{tabular}
\end{center}
$^a$Extrapolated from solution spectra to the gas\textendash phase (see compilation in
Ref.~\cite{kad06}).
\label{hexacene}
\end{table}

\subsection{Quasiparticle\textendash corrected HOMO\textendash LUMO gap of the neutral species}

For each of the neutral molecules considered at the \mbox{B3LYP/6-31+G$^\star$} 
level, we report in Table~\ref{table1} the comparison between the HOMO\textendash LUMO 
gap { E$_\mathrm{gap}^\mathrm{KS}$ obtained as difference of 
Kohn-Sham eigenvalues}, the excitation energy of the HOMO\textendash LUMO transition 
E$_\mathrm{gap}^\mathrm{TD-DFT}$ { as given by TD\textendash DFT}, and the corresponding 
experimental value E$_\mathrm{gap}^\mathrm{exp}$ ($p$\textendash bands in 
Tables~\ref{naphthalene}\textendash\ref{hexacene}). { In the same table we compare 
the results of Eqs.~(\ref{delta}) and (\ref{delta2}) with the DFT\textendash based
tight\textendash binding GW data of Ref.~\cite{nie05}, QP$_\mathrm{gap}^\mathrm{DFT\textendash GW}$,
as well as with the corresponding experimental value  obtained as 
the difference between the experimental EAs and first IEs given in 
Table~\ref{table},
QP$_\mathrm{gap}^\mathrm{exp}$ = IE$_\mathrm{exp}$-EA$_\mathrm{exp}$. 
The theoretical exciton binding energy 
E$_\mathrm{bind}$ is estimated through the difference 
QP$^1_\mathrm{gap}$-E$_\mathrm{gap}^\mathrm{TD-DFT}$,
which is compared with its corresponding experimental value 
QP$_\mathrm{gap}^\mathrm{exp}$ - E$_\mathrm{gap}^\mathrm{exp}$.} All of these 
quantities are displayed in Fig.~\ref{gap} as a function of molecular size.

\begin{landscape}

\begin{table}
\caption{ Comparison between the B3LYP/6\textendash31+G$^\star$ results (all values in 
eV) for the HOMO\textendash LUMO energy gap E$_\mathrm{gap}^\mathrm{KS}$ obtained as 
difference of Kohn\textendash Sham eigenvalues, the excitation energy of the HOMO\textendash LUMO 
transition as given by TD\textendash DFT, {E$_\mathrm{gap}^\mathrm{TD-DFT}$}, and its 
corresponding experimental value E$_\mathrm{gap}^\mathrm{exp}$ ($p$\textendash bands in 
Tables~\ref{naphthalene}\textendash\ref{hexacene}). The QP\textendash gap evaluated using 
Eqs.~(\ref{delta}) and (\ref{delta2}) is compared with the DFT\textendash based tight
binding GW data of Ref.~\cite{nie05},  QP$_\mathrm{gap}^\mathrm{DFT\textendash GW}$, as 
well as with the experimental value QP$_\mathrm{gap}^\mathrm{exp}$ obtained as 
the difference between the experimental EAs and first IEs given in 
Table~\ref{table}. The theoretical and experimental exciton binding energy 
E$_\mathrm{bind}$ are estimated as QP$^1_\mathrm{gap}$-E$_\mathrm{gap}^\mathrm{TD-DFT}$
and QP$_\mathrm{gap}^\mathrm{exp}$ - E$_\mathrm{gap}^\mathrm{exp}$, respectively.
For comparison, the results of Ref.~\cite{kad06} are reported within 
parentheses.}
\begin{center}
\begin{tabular}{c c c c c c c c c c}
\hline \hline
n & E$_\mathrm{gap}^\mathrm{KS}$ & E$_\mathrm{gap}^\mathrm{TD-DFT}$ & 
E$_\mathrm{gap}^\mathrm{exp}$ & 
QP$^1_\mathrm{gap}$ & QP$^2_\mathrm{gap}$ &
{ QP$_\mathrm{gap}^\mathrm{DFT-GW}$}
& QP$^\mathrm{exp}_\mathrm{gap}$  & 
E$_\mathrm{bind}$ & E$_\mathrm{bind}^\mathrm{exp}$\\[2pt]
\hline 
2 & 4.74(4.75) & 4.36(4.35) & 4.45 & 8.27(8.29) & 8.12 & { 8.0} &
8.34 & 3.91 & 3.89\\
3 & 3.54(3.55) & 3.21(3.21) & 3.45 & 6.66(6.60) & 6.58 & { 6.6} &
6.91 & 3.45 & 3.46\\
4 & 2.74(2.75) & 2.45(2.44) & 2.72 & 5.55(5.56) & 5.50 & { 5.5} &
5.90 & 3.10 & 3.18\\ 
5 & 2.19(2.19) & 1.91(1.90) & 2.31 & 4.75(4.76) & 4.72 & { 4.8} &
{ 5.20} & 2.84 & { 2.89}\\
6 & 1.78(1.78) & 1.51(1.50) & 1.90 & 4.15(4.16) & 4.13 &  { 4.3} &
\textemdash{}  & 2.64 & \textemdash\\
\hline
\end{tabular}
\end{center}
\label{table1}
\end{table}

\end{landscape}

\section{Discussion}
\label{discussion}

Figure~\ref{rotconst} shows that structural variations between charged 
species and their respective neutral counterparts display the same 
well\textendash defined trend as a function of molecular size: a general 
decrease for A$_\mathrm{rel}$ (with the possible exception of anthracene anion),
and an increase of B$_\mathrm{rel}$ is observed for all charge\textendash states 
considered. More specifically, the largest structural changes are observed for 
the anions, a consequence of the strongly antibonding character of the LUMO 
of the neutral counterpart. The variations in both cations and dications are 
nearly equal and remain small. A shortening along the short\textendash axis 
(A$_\mathrm{rel}>0$) compared to the neutral molecules is always observed. The 
anions appear to be primarily distorted along the long axis showing a 
lengthening along it (B$_\mathrm{rel}<0$) for all sizes. These 
findings agree with { a} Hartree\textendash Fock study of naphthalene and anthracene 
cations \cite{def93}, and { a} DFT study of anthracene anion performed with 
both hybrid and gradient\textendash corrected exchange\textendash correlation functionals 
\cite{des00}. 

{ Table~\ref{table} and Figure~\ref{ioniz} confirm for the
oligoacenes the good agreement found for the whole class of PAHs between the 
\mbox{B3LYP/6\textendash31+G$^\star$} electron affinities and the available experimental 
data. The differences 
between computed and observed first ionisation energies are of the same order 
of magnitude as in previous analyses \cite{wib97,kad06}, and seem to increase 
sligthly at increasing molecular size (from $\sim$4\% for naphthalene to $\sim$10\% 
for hexacene). 
Double ionisation energies are found to change more rapidly along the series 
than single ionisation energies, and the relative errors of our calculations,
in comparison with the only three experimental data available, increase from 
$\sim$2\% for naphthalene to $\sim$8\% for anthracene and tetracene.}

As shown in Figs.~\ref{anthr} and \ref{benz}, the real\textendash time real\textendash space TD\textendash DFT 
method provides results in very good agreement with the available experimental 
data for neutral species up to about 30 eV. The broad { plasmon\textendash like} 
excitation peaking at about 17.5 eV, which involves $\pi\to\sigma^\star$, $\sigma\to\pi^\star$, $\sigma\to\sigma^\star$,
and Rydberg spectral transitions, is well reproduced both in position and 
width. However, as previously discussed in the case of neutral benzene 
\cite{yab99}, the use of the finite simulation box and the absorbing boundary 
at its edges does not give a satisfactory treatment of continuum effects 
producing spurious structures. The contribution of the three possible 
polarizations to the total absorption cross\textendash section varies considerably. While
the low\textendash energy part is due to in\textendash plane\textendash polarised electronic transitions ($x$ 
and $y$ axes), the contribution of the $z$\textendash axis perpendicular to the plane of 
the molecule is significant only above a few eV. These features are found to be
common to all PAHs \cite{mal}. In the case of oligoacenes, due to their 
special symmetry, the strongest absorption ($\beta$ band in the neutral molecules) 
corresponds to long\textendash axis polarization, which can be simply understood as the 
classical resonance in a conducting rod \cite{pla49}. From Figs.~\ref{comp1} 
and \ref{comp2} it is seen that the broad { plasmon\textendash like} structure with 
its maximum at 17\textendash18 eV is relatively insensitive to the charge\textendash state of the 
molecule. { On the other hand, the charge\textendash state of the molecule} shows up 
in the low\textendash energy range { and,} interestingly, the onset of this broad 
absorption moves blue\textendash ward and becomes steeper with increasing positive 
charge. As shown in Fig.~\ref{integrated} this translates into a systematic 
decrease with increasing positive charge of the absorption cross\textendash section in 
the energy gap between the $\pi$ and $\sigma$ { plasmon\textendash like} structures (between 
$\sim$6 and $\sim$12~eV). This same behaviour has been shown for a larger sample of 
PAHs \cite{mal07a}. In addition, as expected, the observed scatter between the 
different charge\textendash states decreases with increasing molecular size.

Concerning the visible\textendash UV part of the spectrum, as reported in 
Tables~\ref{naphthalene}\textendash\ref{hexacene}, we checked both the sensitivity of 
our TD\textendash DFT calculations to the use of different exchange\textendash correlation 
functionals, and their reliability in comparison with the available 
experimental data. General agreement is found between BLYP and B3LYP which
yield the same ordering of the strongest dipole\textendash allowed excited states. 
{ On the average, the BLYP energies are found to be systematically smaller 
by 0.1\textendash0.6~eV compared to the corresponding B3LYP results }
Our data confirm the spectral assignment done in previous studies for neutral 
\cite{kad06}, and singly\textendash charged species \cite{hir99,hir03}. From the more 
accurate experimental data reported for the anions, neutrals, and cations we 
found that the mean relative deviation of the B3LYP functional is of the order 
of 6\%, compared to the mean relative deviation of about 7\% given by BLYP.
We therefore consider only the former set of results in the following analysis.

The frequency\textendash space implementation of TD\textendash DFT enabled us to gain { some} 
insight into the nature of the first few electronic excitations.  Focusing only
on dications, reported here for the first time, we find that the lowest 
in\textendash plane long ($y$) and short\textendash axis ($x$) polarised bands correspond, 
respectively, to the { \mbox{HOMO-1$\to$LUMO} ($\pi_{-2}\to\pi_1^\star$)} and 
{ \mbox{HOMO-2$\to$LUMO} ($\pi_{-3}\to\pi_1^\star$)} transitions for naphthalene and 
anthracene, and { \mbox{HOMO$\to$LUMO} ($\pi_{-1}\to\pi_1^\star$)} and 
{ \mbox{HOMO-3$\to$LUMO} ($\pi_{-4}\to\pi_1^\star$)} for tetracene, pentacene, and 
hexacene. Figures~\ref{trend1} and \ref{trend2} show interesting trends as to 
the behaviour of the lowest-lying electronic transitions as a function of the 
size of the molecule and its charge state. The positions of both in\textendash plane 
short and long\textendash polarised excitations are found to shift to lower energies with
molecular size for all charge states considered (Fig.~\ref{trend1}). We 
observe the well\textendash known similarities in the electronic absorption spectra 
between anionic and cationic PAHs, as well as the systematic shifts in band 
position when going from the cation to the anion \cite{shi73,hir03}. The sign 
and the magnitude of these shifts, attributed to the different effect of the 
$\sigma$\textendash electrons in both ions \cite{shi73}, are reproduced by TD\textendash DFT for the
most intense bands \cite{hir03}, i.e., the lowest\textendash lying $y$ bands in 
oligoacenes. For example, the measured blue\textendash shifts of 0.07~eV (from 1.43~eV 
of the 1$^2B_{2g}$ state of the cation to 1.50~eV of the 1$^2B_{1u}$ state of the 
anion, data in organic solid \cite{shi73}, see Table~\ref{tetracene}) and 
0.11~eV (from 1.31~eV of the 1$^2A_{u}$ state of the cation to 1.42~eV 1$^2B_{3g}$ 
state of the anion, data in Ne\textendash matrix \cite{hal00}, see 
Table~\ref{pentacene}), when going from the cation to the anion of tetracene 
and pentacene, respectively, compare favorably with our computed blue\textendash shifts 
of 0.07, from 1.70 to 1.77~eV, and 0.10~eV, from 1.50 to 1.60~eV (0.05 and 
0.08 in Ref.~\cite{hir03}). However, this is not the case for the lowest\textendash lying
$x$ bands, for which our theoretical predictions seem to present a mismatch. 
In the case of tetracene and pentacene, e.g., the experimental blue\textendash shifts of 
0.04~eV (from 1.66~eV of the 1$^2B_{3g}$ state to 1.7~eV of the 1$^2A_{u}$ state, 
Table~\ref{tetracene}), and 0.10~eV (from 1.27~eV of the 1$^2B_{1u}$ state to 
1.37~eV of the 1$^2B_{2g}$ state, Table~\ref{pentacene}), are predicted in this 
study to be red\textendash shifted by -0.10, from 1.70 to 1.60~eV, and -0.09~eV, from 
1.25 to 1.16~eV, respectively (-0.06 and -0.05 in Ref.~\cite{hir03}). 

We confirm the system-size-dependent errors found for the short\textendash polarised 
transitions ($p$ bands) in the neutral systems \cite{gri03,kad06}. More 
specifically, from a comparison with the available experimental data reported 
in Tables~\ref{naphthalene}\textendash\ref{hexacene} we find that while the relative 
error in the position of the $\alpha$ band is always of the order of 10\%, in the 
case of the $p$ band this error increases from 2\% for naphthalene to 20\% for 
hexacene (see the { triangles} on the top panel of Fig.~\ref{gap}). On the 
other hand, we find the position of the $\beta$ band to be reproduced with a good 
precision with relative errors of, at most, 4\%. In the case of the radical 
anions and cations, we could not find clear trends as to the performances of 
our TD\textendash DFT/B3LYP results. On the average, the relative errors are larger for 
the long\textendash polarised bands compared to the short\textendash polarised ones for both anions 
(12\% vs. 9\%) and cations (16\% vs. 7\%). 

The oscillator strengths $f$ we obtain for the oligoacenes are also found to 
display systematic changes as a function of molecular size (Fig.~\ref{trend2}).
Unlike { the} neutral molecules, in all charged species the lowest parallel 
($y$) transitions have larger $f$\textendash values compared to the corresponding lowest 
perpendicular ($x$) ones. The oscillator strengths of the $x$\textendash polarised 
transitions display small changes
with the number of benzene units being always of the order of 0.01 for cations 
and anions, and decreasing from 0.06 to 0.03 and from 0.09 to 0.03 for neutrals
and dications, respectively. The oscillator strengths obtained for the 
$y$\textendash polarised transitions are found to increase with the molecular size for 
all the charge states considered. In particular, while the increase for the 
neutral molecules is from $1\cdot10^{-5}$ to $1\cdot10^{-2}$ when going from naphthalene to
hexacene, for the charge states -1, +1, and +2 the corresponding values go from
0.03 to 0.37, 0.05 to 0.32, and from 0.09 to 0.65. Therefore, analogously to 
singly\textendash charged species, doubly\textendash ionised PAHs have strong absorption features in
the near\textendash IR, visible, and near\textendash UV spectral ranges, a result that might be 
relevant in the astrophysical context. 

Orbital energy differences are well\textendash defined zeroth\textendash order approximations to 
electronic excitation energies \cite{gor96}. As shown in Table~\ref{table1}, 
the HOMO\textendash LUMO energy gap obtained directly as difference of Kohn\textendash Sham 
eigenvalues gives a better description of the optical gap, i.e., the position 
of the $p$\textendash band, with relative errors in the range 1\textendash6\%, compared to TD\textendash DFT. 
As already discussed, in this latter case a system\textendash size\textendash dependent error is 
known to exist \cite{gri03}, with an increase from 2 to 20\% in the relative 
error as found in this study { when going from naphthalene to hexacene}.
The { $\Delta$SCF QP\textendash corrected} HOMO\textendash LUMO gaps of neutral oligoacenes as 
computed at the \mbox{B3LYP/6-31+G$^\star$} level, { compare very well with the 
DFT\textendash based tight\textendash binding GW calculated ones independently from the size of the 
system, with discrepancies no larger than 3\%. These values, however, when 
compared to experiments appear to be affected by the same systematic error 
found for the TD\textendash DFT results (see the diamonds on the top panel of 
Fig.~\ref{gap}).} On the other hand, since these errors cancel each other in 
the evaluation of the exciton binding energy, we obtain an accurate estimate 
for E$_\mathrm{bind}$. Appreciable excitonic effects due to both quantum 
confinement and reduction of screening are found, with values ranging from 3.9 
eV for naphthalene to 2.8 eV for hexacene. 

\section{Concluding remarks}
\label{conclusion}

We presented a systematic theoretical study of the five smallest oligoacenes, 
i.e., naphthalene, anthracene, tetracene, pentacene, and hexacene in the
charge states most relevant for astrophysical applications, namely -1, 0, +1,
and +2. From the ground\textendash state structural relaxations performed at the 
\mbox{B3LYP/6-31+G$^\star$} level we computed the electron affinities, the first 
and second ionisation energies, the quasiparticle correction to the HOMO\textendash LUMO 
gap of the neutral systems, and an estimate of the excitonic effects in this 
class of compounds. Good agreement is found with the available experimental 
data as well as with previous theoretical results. To study the electronic 
absorption spectra we used a compendium of the TD\textendash DFT theoretical framework 
in both real\textendash time, to obtain the whole photo\textendash absorption cross\textendash sections in a 
single step, and frequency space, to study general trends as a function of 
charge\textendash state and molecular size for the lowest\textendash lying 
{ valence $\pi\to\pi^\star$} in\textendash plane long\textendash polarised 
and short\textendash polarised electronic transitions. The main step forward achieved in 
this work with respect to previous theoretical analyses lies (i) in the 
spectral range covered, that extends up to the far\textendash UV for both neutral and 
charged PAHs, and (ii) in the first detailed study of doubly\textendash ionised species, 
largely unexplored so far. The interest on PAH dications by the astrophysical
community has been recently renewed
by the proposal that they could be plausible candidates to explain a red 
fluorescence observed in many interstellar sources \cite{wit06}. We find that 
dications, like their singly\textendash charged counterparts, display strong electronic 
transitions of $\pi\to\pi^\star$ character in the near\textendash IR, visible, and near\textendash UV spectral
ranges. As expected, the broad { plasmon\textendash like} structure peaking at about 
17.5~eV is found to be relatively insensitive to the charge\textendash state of the 
molecule, but we interestingly find a systematic decrease with increasing 
positive charge of 
the absorption cross\textendash section between about 6~eV and about 12~eV. Since the 
latter spectral signature is a general property of all PAHs \cite{mal07a}, a 
comparison with astronomical extinction curves could provide an additional 
observational handle for estimating the average charge state of interstellar 
PAHs.

\section*{Acknowledgements}
{ G.~Malloci acknowledges financial support by Regione Autonoma della 
Sardegna. G.~C, G.~M., and G.~M.} acknowledge financial support by MIUR under 
project PON\textendash CyberSar. We thank the authors of \textsc{octopus} for making 
their code available under a free license. We acknowledge the High Performance 
Computational Chemistry Group for the use of \textsc{nwchem}, A Computational 
Chemistry Package for Parallel Computers, Version~4.7 (2005), PNNL,  Richland, 
Washington, USA. Part of the simulations were carried out at CINECA (Bologna).








\end{document}